\documentclass[a4paper,11pt]{article}
\pdfoutput=1

\usepackage{jcappub}
\usepackage{natbib}
\usepackage{aas_macros}
\usepackage{gensymb}
\usepackage{xcolor}
\usepackage{float}

\usepackage{graphics,graphicx, color}
\usepackage{dcolumn, amsmath,amssymb}
\usepackage{booktabs}

\newcommand{\lsim}{\mathrel{\hbox{\rlap{\lower.75ex \hbox{$\sim$}} \kern-.3em \raise.4ex \hbox{$<$}}}}
\newcommand{\gsim}{\mathrel{\hbox{\rlap{\lower.75ex \hbox{$\sim$}} \kern-.3em \raise.4ex \hbox{$>$}}}}

\title{Geometrical Constraints of Observing Very High Energy Earth-Skimming Neutrinos from Space}
               
\author[a]{Claire Gu\'epin}
\author[b]{Fr\'ed\'eric Sarazin}
\author[c,d]{John Krizmanic}
\author[b]{Jacqueline Loerincs}
\author[e]{Angela Olinto}
\author[b]{Ashley Piccone}

\affiliation[a]{Sorbonne Universit\'es, UPMC Univ. Paris 6 et CNRS, UMR 7095,  Institut d'Astrophysique de Paris, 98 bis bd Arago, 75014 Paris, France}
\affiliation[b]{Department of Physics, Colorado School of Mines, 1523 Illinois Street, Golden CO 80401, USA}
\affiliation[c]{University of Maryland, Baltimore County, Baltimore, MD, USA}
\affiliation[d]{CRESST/NASA Goddard Space Flight Center, Greenbelt, MD, USA}
\affiliation[e]{Department of Astronomy \& Astrophysics, KICP, EFI, The University of Chicago, Chicago, IL, USA}

\emailAdd{claire.guepin@iap.fr}

\abstract{The detection of very-high-energy (VHE) neutrinos ($E_\nu \gsim 10$ PeV) is a challenge that future generations of experiments are being designed and constructed to address. One promising method relies on using the Earth as a neutrino target for indirect detection of skimming tau neutrinos interacting within the Earth and producing tau leptons that are able to escape and decay in the atmosphere. The tau decay produces upward-moving Extensive Air Showers (EASs). A space-based or suborbital instrument observing the ground near the Earth limb can search for the beamed Cherenkov signal produced by the up-going EAS resulting from the tau-lepton decay. In this paper, we derive the geometrical constraints for such an observation in general and for the specific case of the Probe Of Extreme Multi-Messenger Astrophysics (POEMMA) mission currently under study, focusing on the Cherenkov signal detection. We show that, using reasonable orbital parameters, POEMMA can achieve full-sky coverage to search for potential neutrino sources over the length of its mission. We also show that follow-up of a transient Target-of-Opportunity (ToO), such as a flaring source, can be achieved within an orbit time scale depending on the source location on the celestial sphere and its relative position with respect to the Sun and the Moon.}

\begin{document}

\renewcommand\d{\mathrm{d}}
\maketitle 

\section{Introduction}

The detailed measurements of the ultra-high-energy cosmic ray (UHECR) spectrum by the Pierre Auger Observatory \citep{Auger} and the Telescope Array experiment \citep{TA}, as well as high-energy (HE) astrophysical neutrinos by the IceCube detector \citep{IceCube13} support the existence of very energetic hadronic accelerators in our universe. In addition to neutrinos from these extreme sources, cosmogenic neutrinos \cite{Berezinsky69} produced by the interaction of UHECRs with the microwave background (the Greisen-Zatsepin-Kuzmin (GZK) effect  \cite{Zatsepin66, Greisen66}) should extend neutrino fluxes beyond the EeV energy scale.  

The very recent correlation of a HE neutrino in coincidence with a Blazar flare \citep{IceCube18} confirms what has been expected for decades, namely that neutrino astronomy will reveal new and unique insights on the highest energy astrophysics in the universe, thus fully opening the multi-messenger window. IceCube has set significant limits on the flux of VHE neutrinos for $E_\nu \lsim 10$ PeV. Auger observations extend these limits above $\sim$ 500  PeV \cite{Zas17ICRC} while the ANITA experiment defines the limits above $\sim$ 30 EeV \cite[][figure 29.10]{Tanabashi18}. These sensitivities only constrain the highest flux levels of the wide range of cosmogenic neutrino fluxes modeled for example by  \cite{Kotera10}. The next generation of UHECR and very-high-energy (VHE, with $E_\nu \gtrsim 10\,{\rm PeV}$) neutrino detectors, such as POEMMA \citep{POEMMA17} for UHECRs and VHE neutrinos, and ARA \citep{Ara12}, ARIANNA \citep{ARIANNA15}, and GRAND \citep{GRAND18} for VHE neutrinos primarily above 100 PeV, are designed to significantly increase UHECR and neutrino sensitivity.  Thus, they will provide critical information about the most energetic messengers received from the universe, with precise measurements of the flux and the composition of the UHECR, as well as the potential first detection of VHE neutrinos through ${\rm EeV}$ energies. In the current context of multi-messenger astronomy, these observations will be decisive to identify the sources of the UHECR and VHE neutrinos and better understand the underlying high-energy astrophysics.

VHE neutrinos are predominately produced by the interactions of VHE- and UHECRs with photons in the acceleration environment of a source (astrophysical) or with photons during their propagation (cosmogenic) \cite{Berezinsky69,HS85,ESS01}. Generated primarily from the decay of charged pions, the produced $\nu_\mu$ and $\nu_e$ spectrum of neutrinos is further mixed due to neutrino oscillations yielding $\nu_e$:$\nu_\mu$:$\nu_\tau$ flavor ratios of 1:1:1. Neutrinos yield precious information about their production sites and their progenitors, as they are not deflected by the extragalactic and galactic magnetic fields during their propagation, and scarcely interact with matter. VHE neutrinos could indeed provide us with unprecedented insight into the cosmic ray acceleration processes, the distribution and evolution of the UHECR sources and the UHECR elementary composition \cite{Kotera10}. In particular, VHE  neutrinos provide a measure of UHECR sources beyond the GZK horizon of $\sim$ 50 Mpc due to the minuscule neutrino interaction cross sections (see \cite{Block14} for a recent calculation).

This also implies that the detection of VHE neutrinos is extremely challenging due to exceedingly long interaction length in a neutrino detection medium. Above about $E_\nu \gtrsim 1\,{\rm PeV}$ however, the neutrino interaction length shortens to a fraction of the Earth radius \cite{Halzen98} making it possible for VHE neutrinos in skimming trajectories to interact inside the Earth \cite{Domokos98a, Domokos98b, Feng02} and allow for the produced tau lepton to escape the Earth before decaying due to the tau lepton's Lorentz-boosted lifetime, modulated by tau energy loss in the Earth.  If the tau lepton decays within the atmosphere, an upward-going extensive air shower will be initiated. The beamed Cherenkov or radio contribution of that shower may be detectable by a space-based instrument viewing the area near the Earth limb. 

This paper investigates the geometrical constraints of detecting the upward Cherenkov component by a detector located in space, including suborbital altitudes. The general characteristics of a space-based detector dedicated to the detection of VHE Earth-skimming tau neutrinos are described in Section~\ref{Sec:general}, as well as the geometrical approach used to calculate the effective field of view and the sky exposure related to the observation of VHE tau neutrinos. In Section~\ref{Sec:Nu_fov}, we study the influence of several parameters, such as the altitude of the detector and the viewing angle, and of the focal surface design specific to optical Cherenkov detection for a configuration specific to the POEMMA Schmidt telescopes, and present estimates of the sky exposure. The question of the sky coverage is addressed in Section~\ref{Sec:full_sky}. Finally, the possibility of Target-of-Opportunity (ToO) observations is addressed in Section~\ref{Sec:Nu_ToO}, where we evaluate the possibility of observing any direction of the sky in a short timescale. Our approach focuses mainly on geometrical calculations, therefore we did not include a consideration of cloud coverage. Moreover, a limited comprehensive study of the influence of the Sun and the Moon on the observations was included. The effects of clouds should be incorporated in a more detailed study as they will influence the sky exposure and sky coverage calculations.

\section{General characteristics of space-based VHE neutrino detector}\label{Sec:general}

\begin{figure}[ht]
\centering
\includegraphics[width=0.5\textwidth]{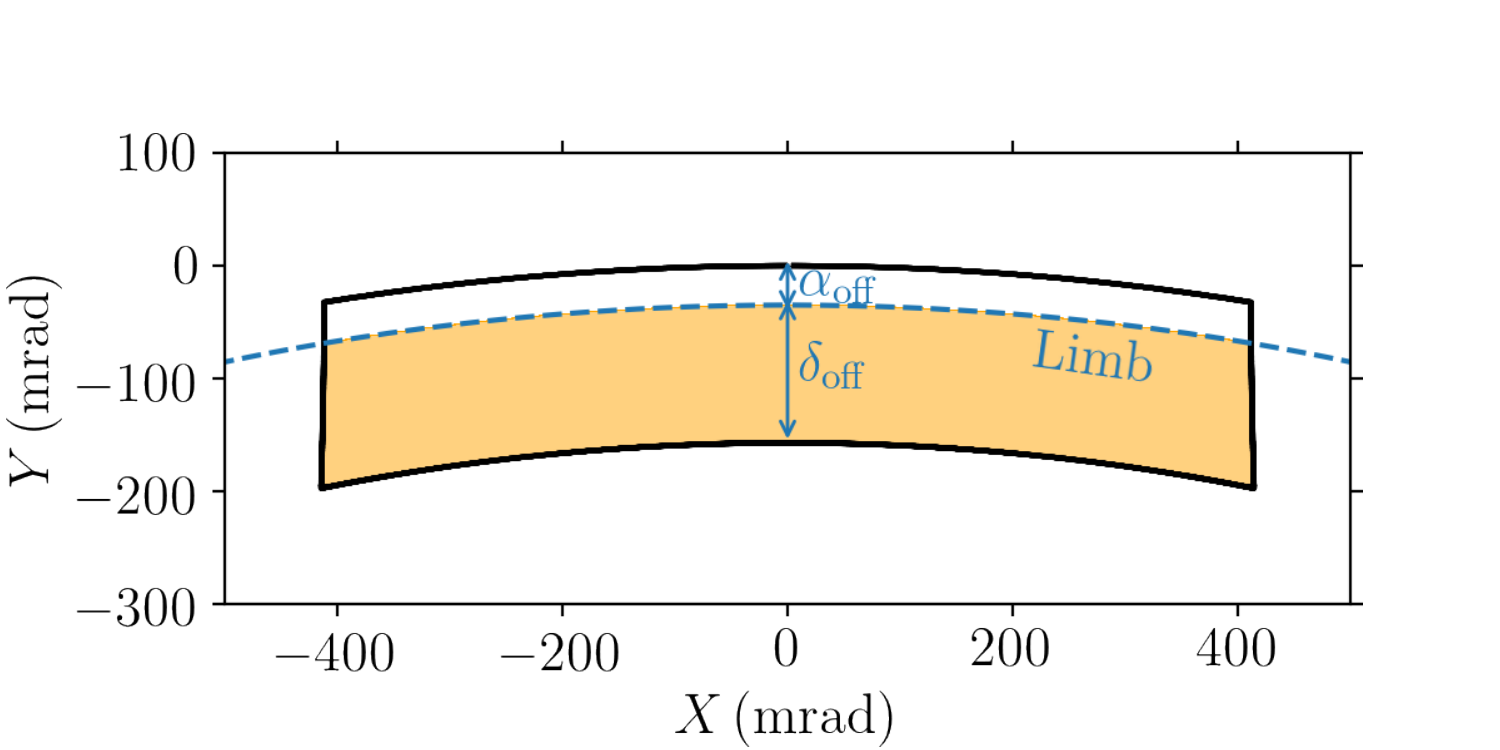}
\caption{Illustration of a generic focal surface for the indirect detection of VHE Earth-skimming tau neutrinos, using the typical opening angle around $45\degree / 2$. The limb of the Earth is shown as a dashed blue line. The orange area corresponds to the part of the focal surface below the limb, that can detect VHE Earth-skimming tau neutrinos.}
\label{Fig:FOV_general_1}
\end{figure}

The basic field-of-view (FOV) characteristics of a space instrument dedicated to the detection of the Cherenkov light arising from a VHE tau neutrino interaction are naturally constrained. The detector needs to point at the Earth limb with the widest possible angular coverage, especially in azimuth (one could envision several individual units mounted in the same satellite like in the CHANT proposal \cite{CHANT}). Its coverage should extend below the limb until the probability of an escaping  tau lepton becomes marginal for all detectable neutrino energies. Typically, the instrument should also cover a few degrees above the limb to allow for background estimates, including air glow and UHECR-induced Cherenkov signals. Fig.~\ref{Fig:FOV_general_1} shows an example of what the FOV angular coverage of a dedicated generic instrument may look like. In this example, the instrument has a FOV angle $\theta_{\rm FOV} = 45\degree$ and covers a region ranging from $\delta=7\degree$ below the limb to $\alpha_{\rm off}=2\degree$ above the limb. We note that in practice, it is preferable that the detector is not oriented towards the satellite moving (ram) direction to prevent potential degradation of the optics (unless properly coated) of the Cherenkov detection instrument due to the atomic oxygen present in the upper atmosphere \cite{Banks04}.

\begin{figure}[ht]
\centering
\includegraphics[width=0.5\textwidth]{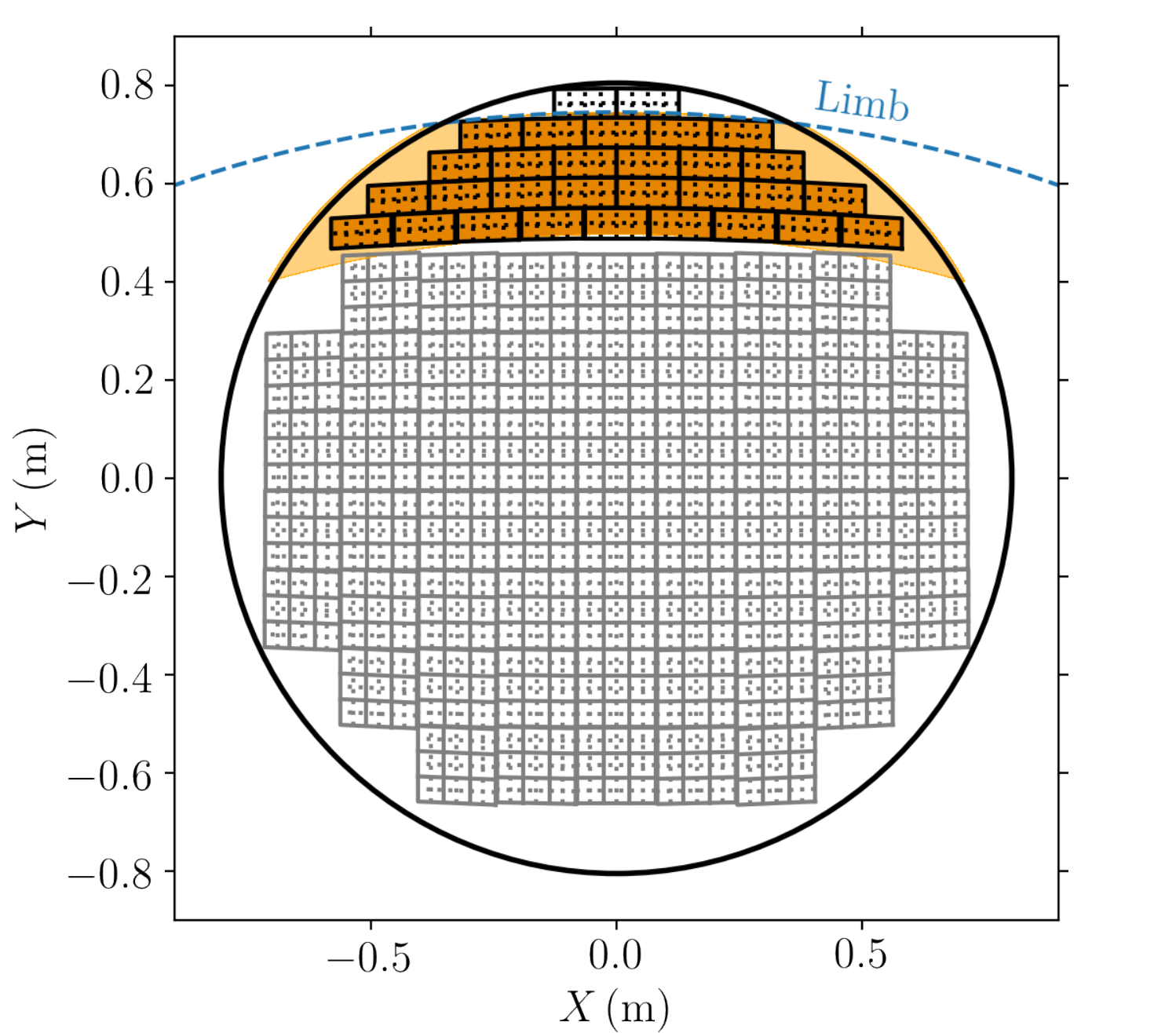}
\caption{Focal surface of the POEMMA detector with the photosensors dedicated to VHE neutrino detection (black tiles) and to the UHECR detection (grey tiles). The projection of the geometrical FOV on the focal surface is shown in light orange and the fraction that can be detected by the tiles is shown in darker orange. The limb of the Earth is shown as a dashed blue line. }
\label{Fig:focal_surf_1}
\end{figure}

However, in what follows we will consider a different FOV configuration. POEMMA is conceived as a system of two satellites in trailing orbits, which aims at detecting both UHECR and VHE neutrinos in a dual-observation mode. The design of the instrument revolves around a uniform 45$\degree$ FOV equipped with two types of photo-detectors. The top of the FOV, dedicated to the detection of VHE neutrinos, uses Silicon photomultiplier (SiPM) detectors and matches the shape of the dedicated instrument shown in Fig.~\ref{Fig:FOV_general_1}. The bottom of the FOV,  dedicated to the detection of UHECRs, uses an arrangement of Multi-Anode-PhotoMultiplier-Tubes (MAPMTs). With the top of the FOV observing the limb, the bottom of the FOV is looking down at the Earth surface, where UHECR showers can be observed. This is what makes the dual purpose FOV of POEMMA so attractive. Fig.~\ref{Fig:focal_surf_1} shows the specific angular coverage of POEMMA for VHE neutrinos with $\delta_{\rm off}=7\degree$ and $\alpha_{\rm off}=2\degree$. As can be seen, the Cherenkov-sensitive part of the POEMMA focal surface follows the same philosophy as the generic detector shown in Fig.~\ref{Fig:FOV_general_1}.

\begin{figure}[ht]
\centering
\includegraphics[width=0.54\textwidth]{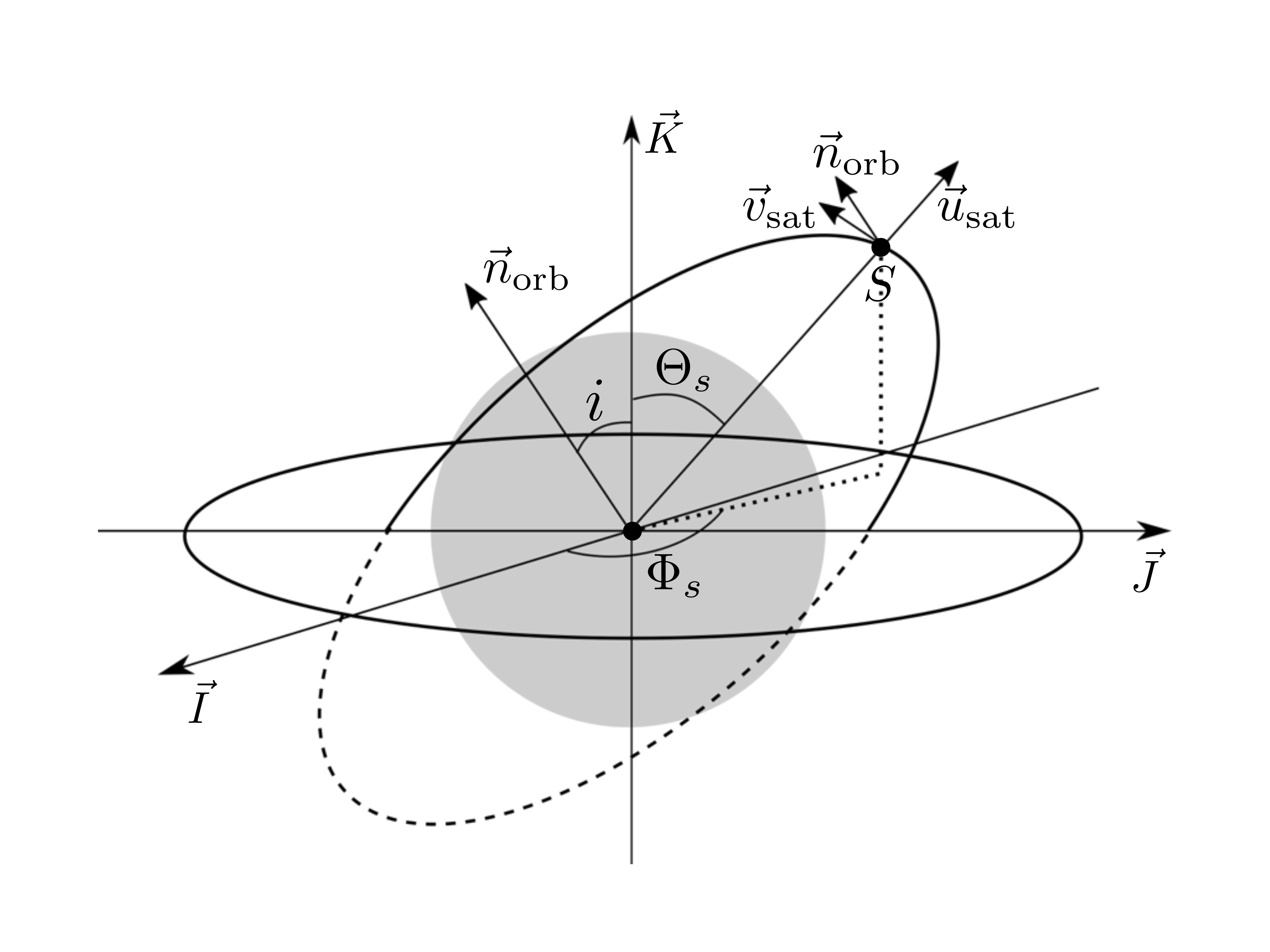}
\includegraphics[width=0.45\textwidth]{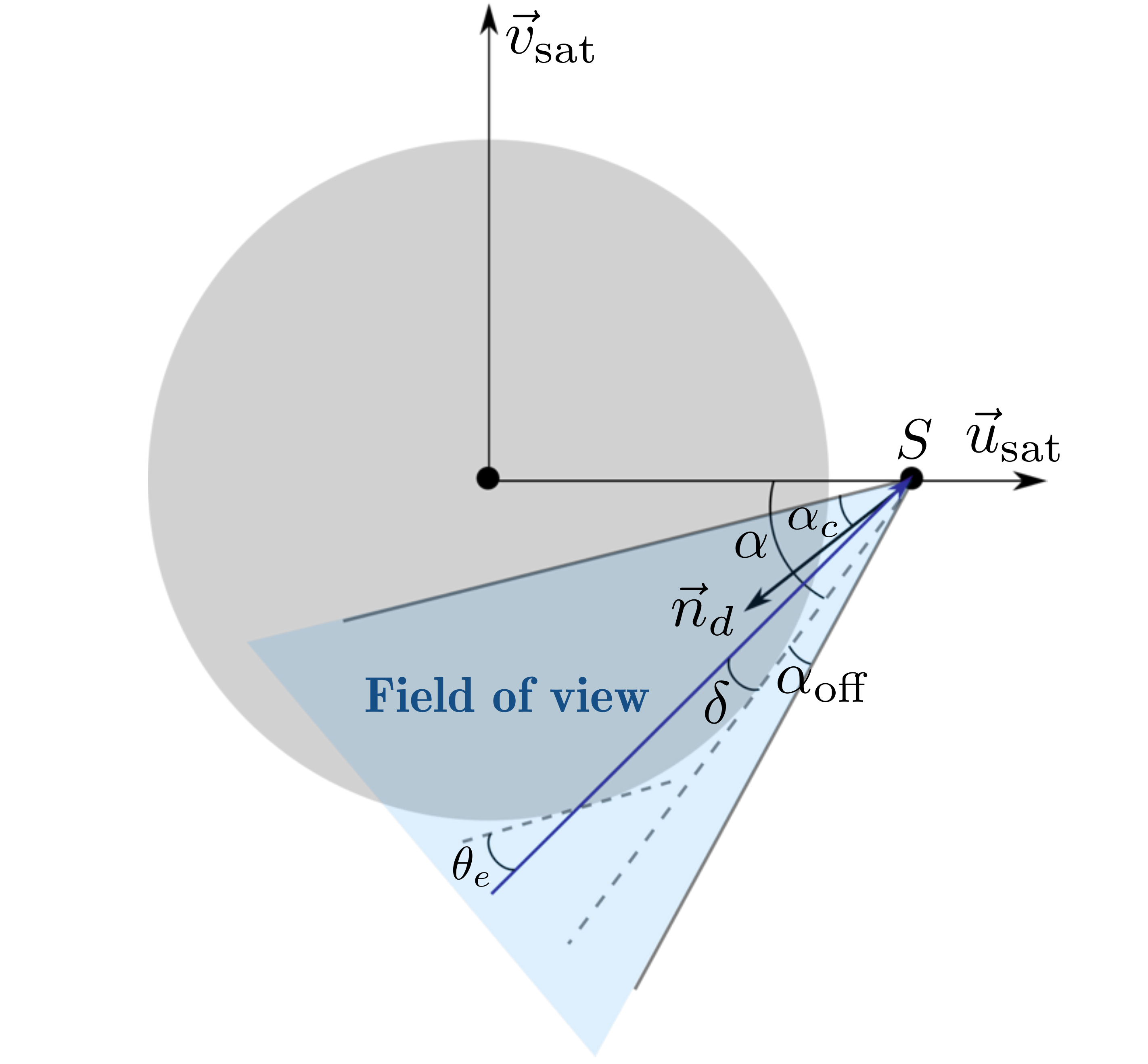}
\caption{Orbital characteristics and notations (left) and illustration of the geometrical configuration in the orbital plane $(\vec{u}_{\rm sat},\vec{v}_{\rm sat})$ (right). The satellite is located at point $S$. A neutrino arriving at the detector is characterized by its Earth emergence angle $\theta_e$ and the corresponding viewing angle $\delta$ from the satellite point of view. The detector has a conical FOV of opening angle $\alpha_c$, with an offset angle $\alpha_{\rm off}$ (away from the Earth limb) and pointing direction $\vec{n}_d$.}\label{Fig:Nu_obs}
\end{figure}

To carry out our calculations, we consider one satellite on a circular orbit of inclination angle $i$, located at an altitude $h$.  The configuration and the notations are illustrated in Fig.~\ref{Fig:Nu_obs}, in the equatorial coordinate system, with $\vec{I}$ the vernal equinox and $\vec{K}$ the north pole. At a given time, the satellite is characterized by its position $(\Theta_s,\Phi_s)$\footnote{$\Theta_s = \pi/2 - \delta_s$ and $\Phi_s = \alpha_s$ with $\alpha_s$ and $\delta_s$ the right ascension and declination of the satellite.}, with $\vec{u}_{\rm sat} = \sin \Theta_s \cos \Phi_s \vec{I} + \sin \Theta_s \sin \Phi_s \vec{J} + \cos \Theta_s \vec{K}$,  such as $\vec{n}_{\rm orb} \cdot \vec{u}_{\rm sat}  = 0$. For instance for $\vec{n}_{\rm orb}$ in the $(\vec{K},\vec{I})$ plane, $\vec{n}_{\rm orb} = \cos i \,\vec{K} + \sin i \,\vec{I}$ and we obtain the constraint $\cos i \cos \Theta_s + \sin i \sin \Theta_s \cos \Phi_s = 0$. The period of the orbit is given by $P=2 \pi \sqrt{(R_\oplus + h)^3/(G M_\oplus)}$ where $G$ is the gravitational constant and $M_\oplus$ the mass of the Earth; thus for instance $P\approx1\,{\rm h}\,45\,{\rm min}$ for $h=1000\,{\rm km}$ and $P\approx1\,{\rm h}\,35\,{\rm min}$ for $h=525\,{\rm km}$. Moreover the orbit precesses around the north pole, with precession rate $\omega_p = (3/2) R_\oplus^2/(R_\oplus+h)^2 J_2 \,\omega \cos(i)$, where $\omega=2\pi/P$ and $J_2$ is related to the Earth oblateness. For an orbit inclination $i=28.5\degree$, the orbit precesses in $P_p\approx68\,{\rm d}\;13\,{\rm h}\;39\,{\rm min}$ for $h=1000\,{\rm km}$ and in $P_p\approx54\,{\rm d}\;7\,{\rm h}\;26\,{\rm min}$ for $h=525\,{\rm km}$.

The configuration in the plane containing the vectors $\vec{u}_{\rm sat}$ and $\vec{n}_{\rm d}$, where $\vec{n}_{\rm d}$ is the pointing axis of the detector, is also illustrated in Fig.~\ref{Fig:Nu_obs}. The fraction of the Earth crossed by the tau neutrino and the subsequent tau lepton is characterized by its Earth emergence angle $\theta_e$ (where $\pi/2-\theta_e$ is the local zenith angle), or the equivalent viewing angle $\delta = \arcsin\left[R_\oplus/(R_\oplus+h) \right] - \arcsin\left[R_\oplus/(R_\oplus+h) \sin(\pi/2-\theta_{e})\right] $. A comprehensive calculation of the propagation and interaction chain of the tau neutrino and the tau lepton can be found in \cite{Reno18} and \cite{Alvarez17}. As the produced Cherenkov signal is significantly beamed, it is detectable only if the arrival direction points toward the detector, within the FOV determined by the EAS Cherenkov angle, which is $\sim 1.4^\circ$ in air at standard temperature and pressure (STP). In the following, we consider the maximum emergence angle $\theta_e$, above which the probability that a $\tau$ lepton escapes the Earth is very small, even considering regeneration effects \cite{Alvarez17,Reno18}. Several estimates of the maximum emergence angles have been made for various neutrino energies. From \cite{Alvarez17}, $\theta_e=5\degree$ appears as a good approximation for $E_\nu = 10^{18}\,{\rm eV}$, as for larger angles the probability that the tau lepton exits the Earth drops. However, these results depend on various factors, for instance the composition of the materials the tau lepton traverses or the neutrino energy. As suggested by recent calculations, we can also consider larger emergence angles and, in the following, the values $\theta_e=14.4\degree$ and $\theta_e=19.6\degree$ (respectively $\delta = 4\degree$ and $\delta = 7\degree$) are also often used and corresponds to VHE tau neutrino energies down to $\sim 10$ PeV \cite{Reno18}.

In order to determine, for a specific satellite orbital position, the portion of the sky detectable in neutrinos, subsequently called effective or geometrical FOV, we consider two conditions for tau neutrino detectability. The first condition is related to the physical properties of tau neutrino and tau lepton interactions, and thus involves the column depth of Earth crossed or the maximum emergence angle $\theta_e$. The second condition is related to the characteristics of the detector, and thus concerns the arrival direction of the Cherenkov signal (and therefore the arrival direction of the tau neutrino), which should be within the FOV. Note that at VHE energies, the direction of the tau and neutrino are well aligned. We note that only this second condition should be adjusted in the case of a different detector design. The first condition is fulfilled if \begin{equation}\label{Eq:cond_emangle}
\arcsin\left[\frac{R_\oplus}{R_\oplus+h} \sin(\pi/2-\theta_{e})\right]< \arccos (\vec{u} \cdot (-\vec{u}_{\rm sat}) ) < \alpha
\end{equation}
where $\alpha = \arcsin[R_\oplus/(R_\oplus+h)]$, $h$ is the altitude of the satellite and $R_\oplus$ the Earth radius. The angle between the source direction $\vec{u}$ and nadir $-\vec{u}_{\rm sat}$ is indeed constrained by the maximum emergence angle (lower bound) and the Earth limb (upper bound). For a conical FOV, the second condition is fulfilled if \begin{equation} \vec{u} \cdot \vec{n}_d > \cos \alpha_c \, , \end{equation}
where $\vec{n}_d$ is the detector axis and $\alpha_c = \theta_{\rm FOV}/2$ is the opening angle of the detection cone. In this case, the effective FOV is thus the intersection between a cone of opening angle $\alpha_c$ and a conical shell.

Simplified estimates of the sky exposure can be obtained by following the approach presented in \cite{Casadei05}, which is valid only if the eccentricity of the orbit is zero, the orbit precession period is much larger than the orbital period, and the detector FOV is a cone whose axis lies in the orbital plane and rotates at the satellite orbital angular velocity. Whereas the first two points are in general valid, the last assumption is not true for all orbital configurations. Indeed, the detector is likely to point off orbit, for instance in order to achieve full-sky coverage or meet requirements concerning a specific observation mode (see for instance Sections~\ref{Sec:full_sky} and \ref{Sec:Nu_ToO}). Alternatively, the effective FOV might not be a cone, and have a more complex structure due to detection constraints: for instance, the interaction of neutrinos with the Earth induces geometrical constraints on the arrival direction of the signal and thus on the detection. Therefore, for any orientation of the detector, we perform more realistic estimates of the portion of the sky available for VHE tau neutrino observation by integrating the exposure time over the observation time for a geometrically constrained FOV.

For a small satellite movement along its orbit, from $\Phi_s$ to $\Phi_s+\Delta\Phi_s$, the detector observes approximately the same portion of the sky (for instance the one observable at $\Phi_s$), and the exposure is related to the time of displacement $\Delta t = t(\Phi_s+\Delta\Phi_s)-t(\Phi_s)$. We consider the equatorial plane ($\vec{I},\vec{J},\vec{K}$) as the reference frame (see Fig \ref{Fig:Nu_obs}). The non-rotating orbital plane frame is ($\vec{i}_{\rm orb},\vec{j}_{\rm orb},\vec{n}_{\rm orb}$). We focus on the case where $\vec{j}_{\rm orb}=\vec{J}$, thus $\vec{i}_{\rm orb}= -\cos i \vec{I} + \sin i \vec{K}$ and $\vec{n}_{\rm orb}= \sin i \vec{I} + \cos i \vec{K}$, where $i$ is the inclination of the orbit. For a position $S$ of the satellite, with an angle $\alpha_{\rm sat}$ in the orbital plane, such as $\vec{u}_{\rm sat} \equiv \cos \alpha_{\rm sat} \vec{i}_{\rm orb} + \sin \alpha_{\rm sat} \vec{J} = \sin \Theta_s \cos \Phi_s \vec{I} + \sin \Theta_s \sin \Phi_s \vec{J}  + \cos \Theta_s \vec{K}$, and projecting on $\vec{I}, \vec{J}$ and $\vec{K}$, we obtain:
\begin{eqnarray}
&&\vec{u}_{\rm sat} \cdot \vec{I} = - \cos \alpha_{\rm sat} \cos i = \sin \Theta_s \cos \Phi_s  \, , \nonumber\\
&&\vec{u}_{\rm sat} \cdot \vec{J} = \sin \alpha_{\rm sat} = \sin \Theta_s \sin \Phi_s \, ,\nonumber\\
&&\vec{u}_{\rm sat} \cdot \vec{K} = \cos \alpha_{\rm sat} \sin i = \cos \Theta_s \, .\nonumber
\end{eqnarray}
If $\cos \Theta_s \neq 0$, we have $ \alpha_{\rm sat} = \arctan \left( -\cos i \tan \Phi_s \right)$ and we obtain the time of displacement 
\begin{equation}\label{Eq:time}
\Delta t = \frac{1}{\omega} \left| \arctan\left[ -\cos i \tan (\Phi_s+\Delta\Phi_s) \right] -\arctan\left[ -\cos i \tan \Phi_s \right] \right|
\end{equation}
where $\omega$ is the angular speed of the satellite. One needs also to consider some particular cases, for instance when $\cos \Phi_s=0$ or $\sin \Phi_s=0$. The above formula allows to calculate the exposure time for any observation time, as a function of the right ascension and the declination, or integrated over the celestial sphere.

\section{Effective field of view and sky exposure estimates}\label{Sec:Nu_fov}

In this section, we illustrate the method described above to calculate the effective FOV and the sky exposure, in the case of a conical FOV, using the characteristics of the POEMMA instrument. As a first step, we consider that the detector is pointing in the direction opposite to the moving direction, with the detector axis in the orbital plane. In the case of POEMMA, we are only considering the FOV of one satellite as both satellites are in close proximity and observing in the same direction. The pointing direction is chosen such that the edge of the detection cone is above the limb of the Earth with an offset $\alpha_{\rm off}$, defined as positive when above the Earth limb, thus $\vec{n}_d = -[\cos (\alpha-\alpha_c+\alpha_{\rm off}) \,\vec{u}_{\rm sat} + \sin (\alpha-\alpha_c+\alpha_{\rm off}) \,\vec{v}_{\rm sat}]$. In the following, we calculate the solid angle on the celestial sphere $\Omega$ corresponding to the part of the sky observable in neutrinos, which allows to study the impact of several parameters, and to give an example of a focal surface specific to a POEMMA instrument.

\subsection{Influence of emergence angle, offset angle and altitude}

\begin{figure}[ht]
\centering
\includegraphics[width=\textwidth]{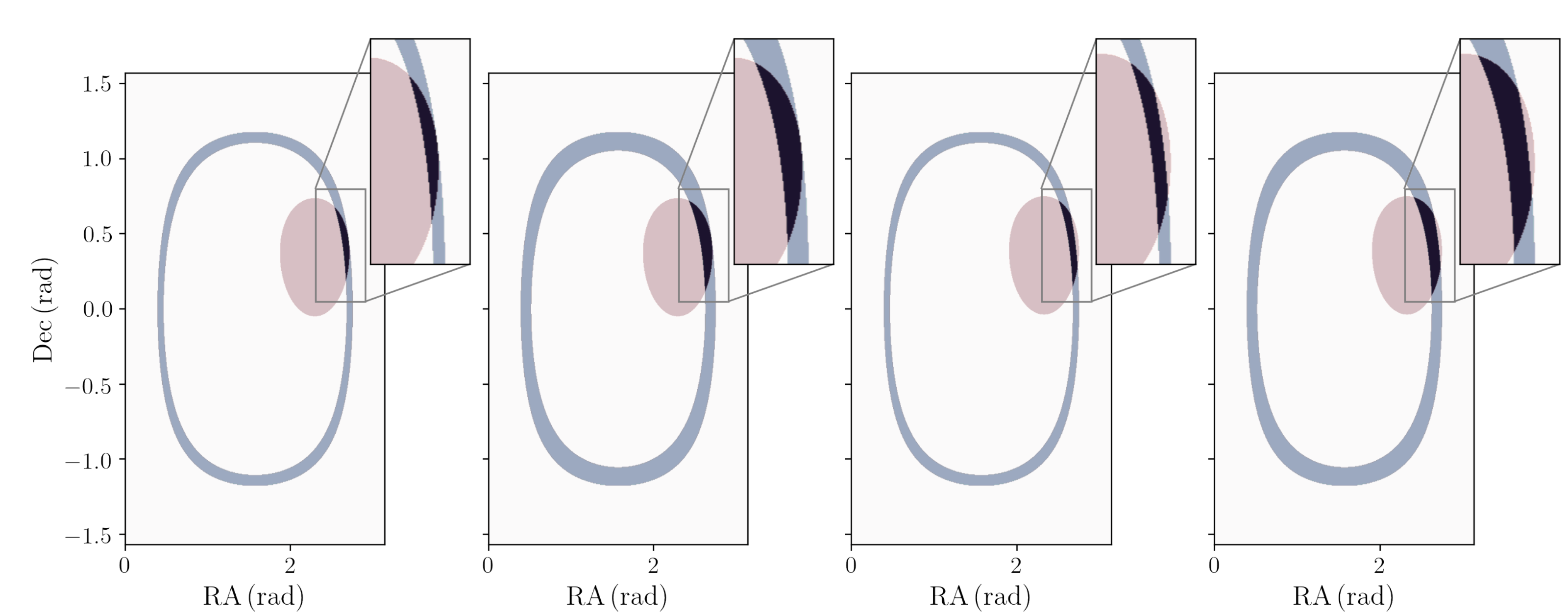}
\caption{Observable portion of the sky as a function of right ascension (RA) and declination (Dec) for a given satellite position at altitude $h=525\,{\rm km}$, with $\theta_e=14.4\degree$ and $\theta_e=19.6\degree$, for $\alpha_{\rm off} = 0\degree$, and $\alpha_{\rm off} = 2\degree$ (from left to right). The red ellipse is related to the condition on the FOV, the blue band to the condition on the emergence angle and the common fraction is shown in black.}\label{Fig:Nu_sky}
\end{figure}

\begin{figure}[ht]
\centering
\includegraphics[width=0.35\textwidth]{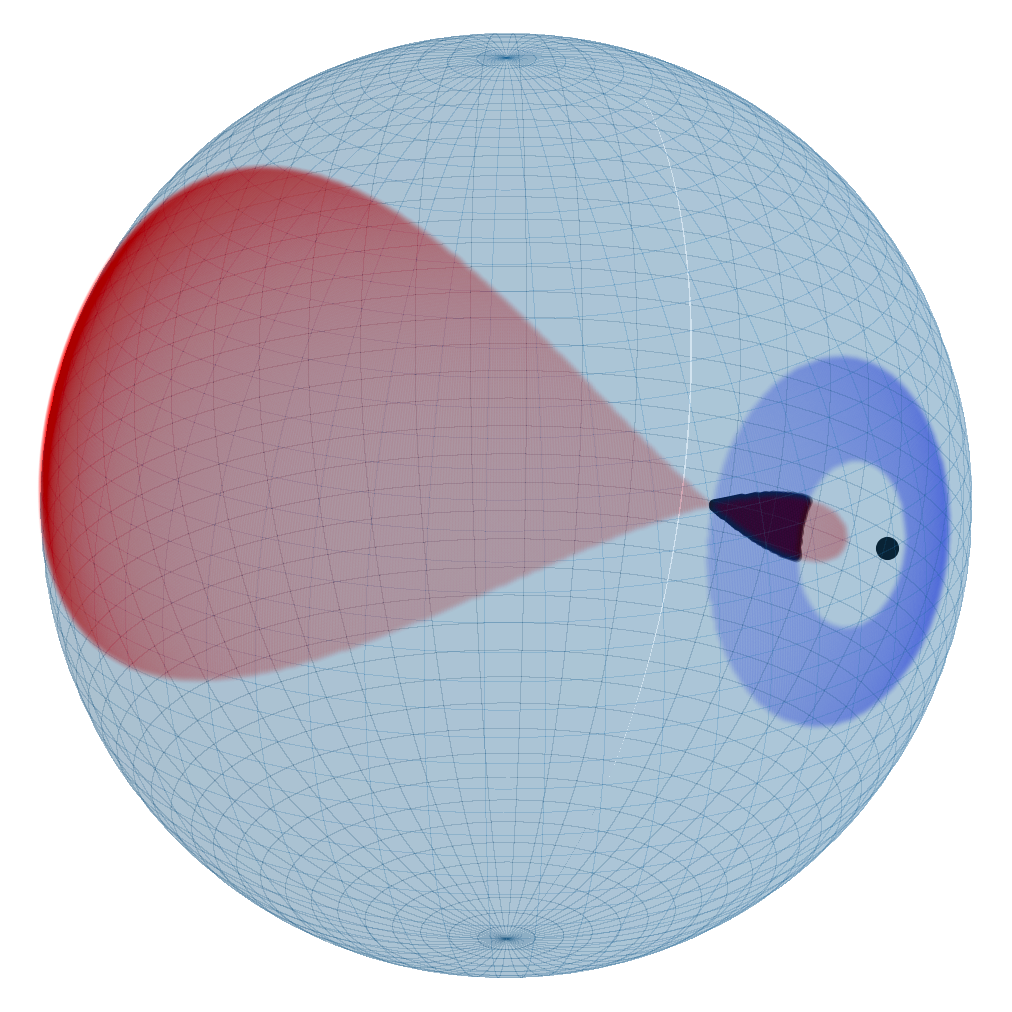}
\hspace{1cm}
\includegraphics[width=0.35\textwidth]{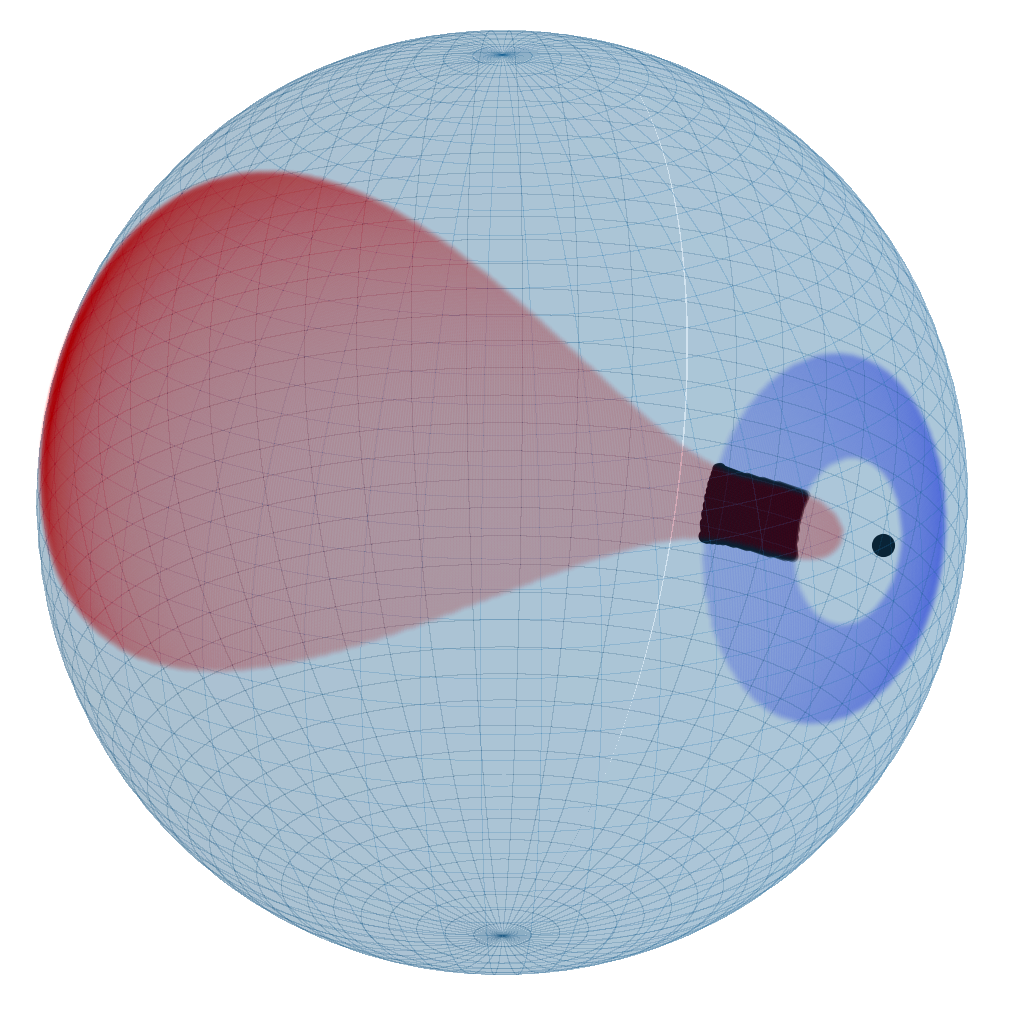}
\caption{Effective FOV projected on the Earth surface, for one satellite position, $\theta_e=19.6\degree$ and $h=525\,{\rm km}$, for $\alpha_{\rm off} = 0\degree$ (left) and $\alpha_{\rm off} = 2\degree$ (right). The red part is related to the condition on the FOV, the blue part to the condition on the emergence angle and the common area is shown in black.}\label{Fig:Nu_area}
\end{figure}

Several estimates of the solid angle $\Omega$ are shown in Table~\ref{Tab:Nu_results_comp}, for a given satellite position at two altitudes $h=525\,{\rm km}$ and $h=1000\,{\rm km}$, for different emergence angles $\theta_e$ and offset angles $\alpha_{\rm off}$.  We note that these estimates varies little along the orbit, thus the following estimates can be considered as typical values. As an example, for $\theta_e=19.6\degree$ and $\alpha_{\rm off}=2\degree$, the solid angle varies along one orbit between $\Omega = 6.7 \times 10^{-2}\,{\rm sr}$ and $\Omega = 7.1 \times 10^{-2}\,{\rm sr}$ for $h=525\,{\rm km}$, and between $\Omega = 4.9 \times 10^{-2}\,{\rm sr}$ and $\Omega = 5.3 \times 10^{-2}\,{\rm sr}$ for $h=1000\,{\rm km}$. The observable fraction of sky projected on the celestial sphere (equatorial coordinates) is illustrated in Fig.~\ref{Fig:Nu_sky} for a given satellite position at altitude $h=525\,{\rm km}$, the emergence angles  $\theta_e=14.4\degree$ and $\theta_e=19.6\degree$ and the offset angles $\alpha_{\rm off}=0\degree$ and $\alpha_{\rm off}=2\degree$. The projection on the Earth surface is also illustrated in Fig.~\ref{Fig:Nu_area} for a given satellite position at altitude $h=525\,{\rm km}$, an emergence angle $\theta_e=19.6\degree$ and the offset angles $\alpha_{\rm off}=0\degree$ and $\alpha_{\rm off}=2\degree$.

\begin{table}[h]
\begin{center}
\begin{tabular}{lccc}
& \begin{tabular}{rr} \multicolumn{2}{c}{$\Omega$ [sr]}  \end{tabular} & \begin{tabular}{rr} \multicolumn{2}{c}{$\Omega$ [sr]}  \end{tabular}\\
\toprule & \begin{tabular}{rr} \multicolumn{2}{c}{$h=525\,{\rm km}$}  \end{tabular} & \begin{tabular}{rr} \multicolumn{2}{c}{$h=1000\,{\rm km}$}  \end{tabular} \\
\midrule & \begin{tabular}{rr} $\alpha_{\rm off} = 0\degree$ & $\alpha_{\rm off} = 2\degree$ \end{tabular}  &  \begin{tabular}{rr} $\alpha_{\rm off} = 0\degree$ & $\alpha_{\rm off} = 2\degree$ \end{tabular} \\
\midrule $\theta_e = 5\degree$ &  \begin{tabular}{rr} $1.2\times 10^{-3}$ & $3.4\times 10^{-3}$ \end{tabular}  &  \begin{tabular}{rr} $7.7\times 10^{-4}$ & $2.5\times 10^{-3}$ \end{tabular}   \\
\midrule  $\theta_e = 14.4\degree$ &  \begin{tabular}{rr} $2.4\times 10^{-2}$ & $3.4\times 10^{-2}$ \end{tabular}  &  \begin{tabular}{rr} $1.6\times 10^{-2}$ & $2.5\times 10^{-2}$ \end{tabular}   \\
\midrule  $\theta_e = 19.6\degree$ &  \begin{tabular}{rr} $5.4\times 10^{-2}$ & $6.7\times 10^{-2}$ \end{tabular}  &  \begin{tabular}{rr} $3.7\times 10^{-2}$ & $5.0\times 10^{-2}$ \end{tabular}   \\
\bottomrule
\end{tabular}
\caption{Solid angles for different sets of emergence angles, offset angles and satellite altitudes.}
\label{Tab:Nu_results_comp}
\end{center}
\end{table}

We note that the blue region in Fig.~\ref{Fig:Nu_sky} and Fig.~\ref{Fig:Nu_area} gets thickers: as expected, the observable fraction of the sky increases with increasing emergence angle. Similarly, the observable fraction of the sky increases with the FOV angle, which corresponds to an increase of the size of the red region in Fig.~\ref{Fig:Nu_sky} and Fig.~\ref{Fig:Nu_area}. We note that in Fig.~\ref{Fig:Nu_area}, a large fraction of the red area, located outside the outer part of the blue region, lies below the limb, and is therefore not accessible to detection. The addition of an offset angle only influences the condition on the FOV, as can be seen in Fig.~\ref{Fig:Nu_sky} and Fig.~\ref{Fig:Nu_area}. The observable portion of the sky should increase by increasing the offset angle from $0$ to approximately $\alpha_c$ and then decrease. Depending on the characteristics of the detector, such as the constraints on the focal surface, one can maximize the observable fraction of the sky by choosing the optimal offset angle (assuming a conical FOV).

\begin{figure}[ht]
\centering
\includegraphics[width=0.48\textwidth]{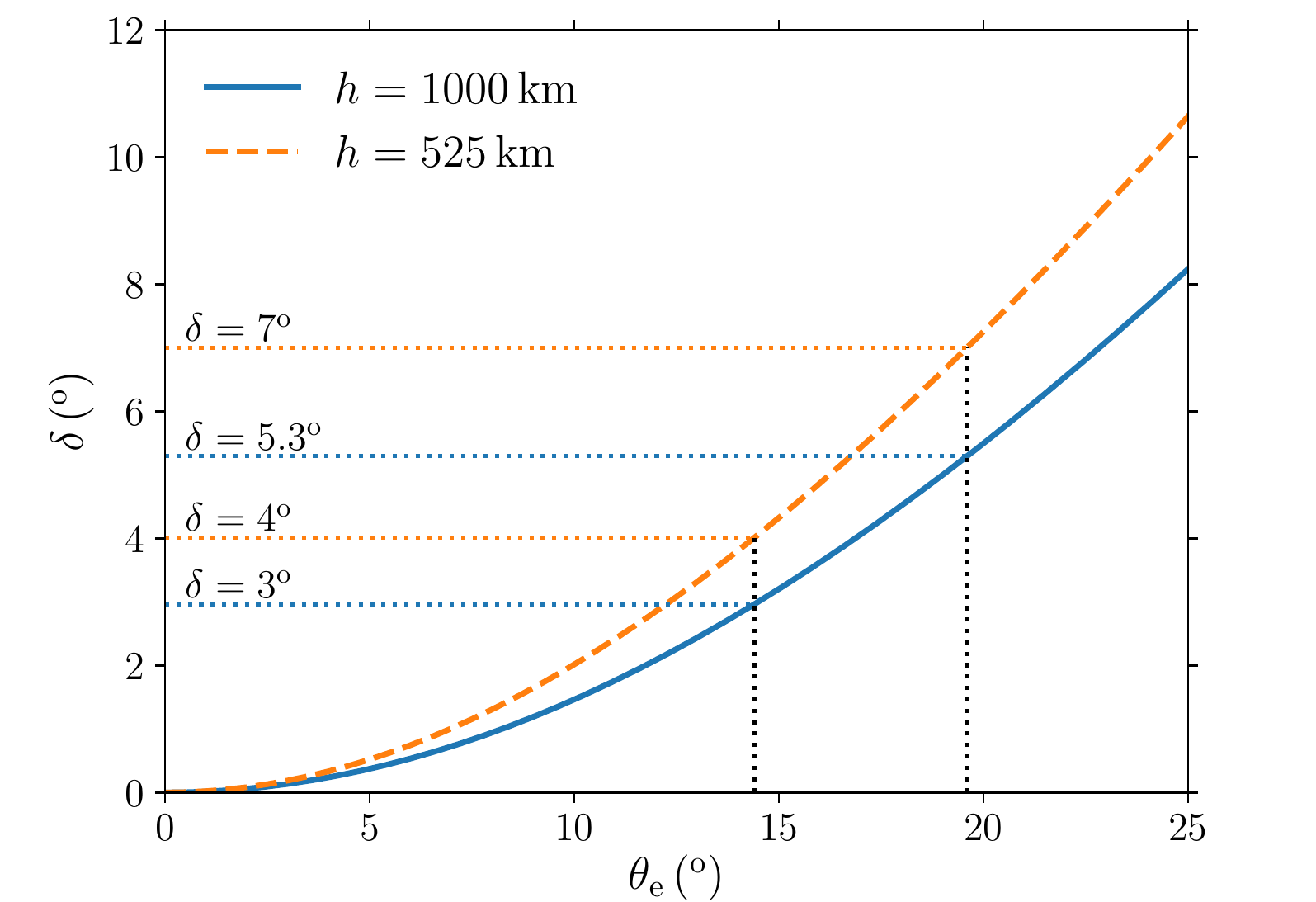}\\
\includegraphics[width=0.49\textwidth]{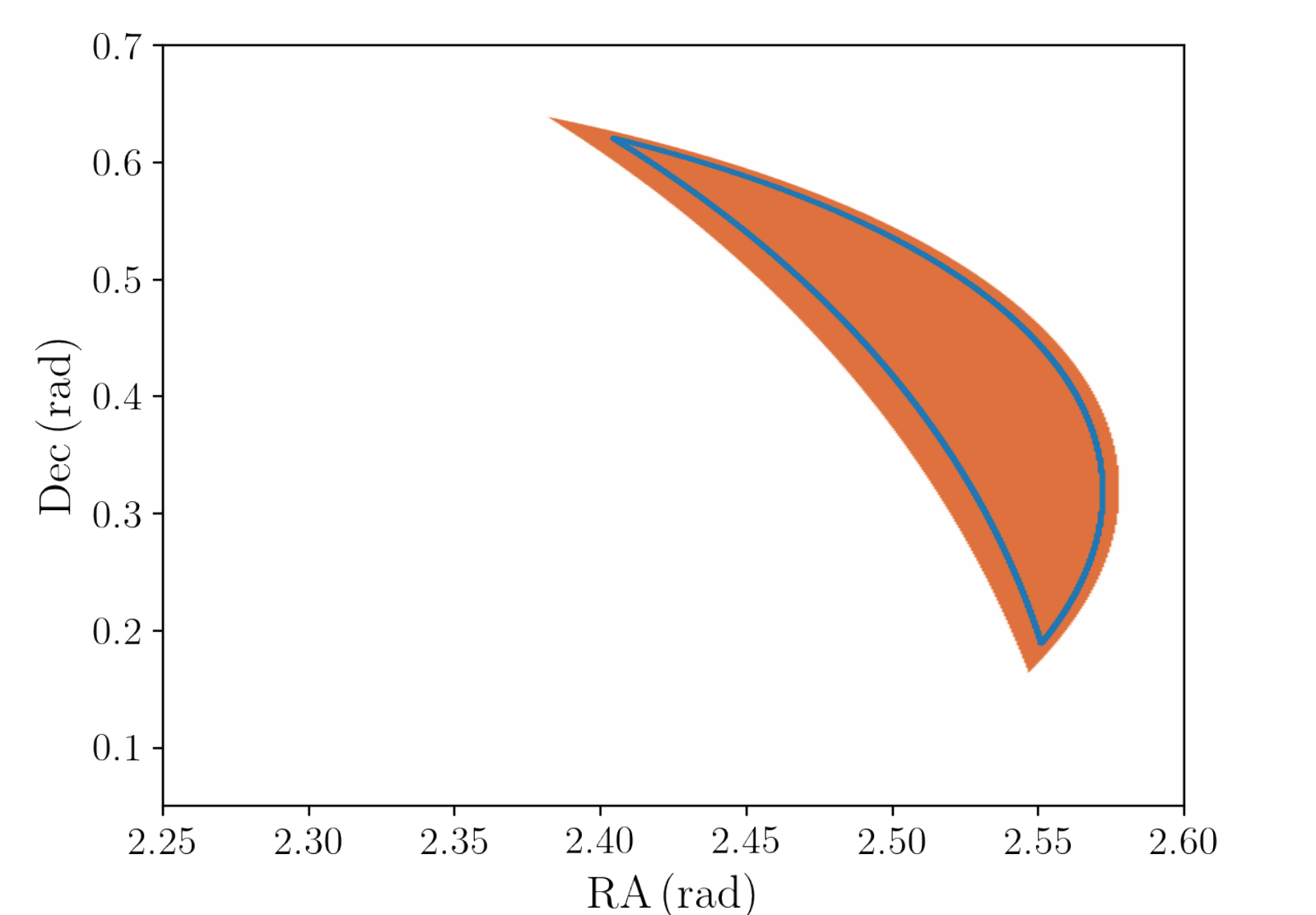}
\includegraphics[width=0.49\textwidth]{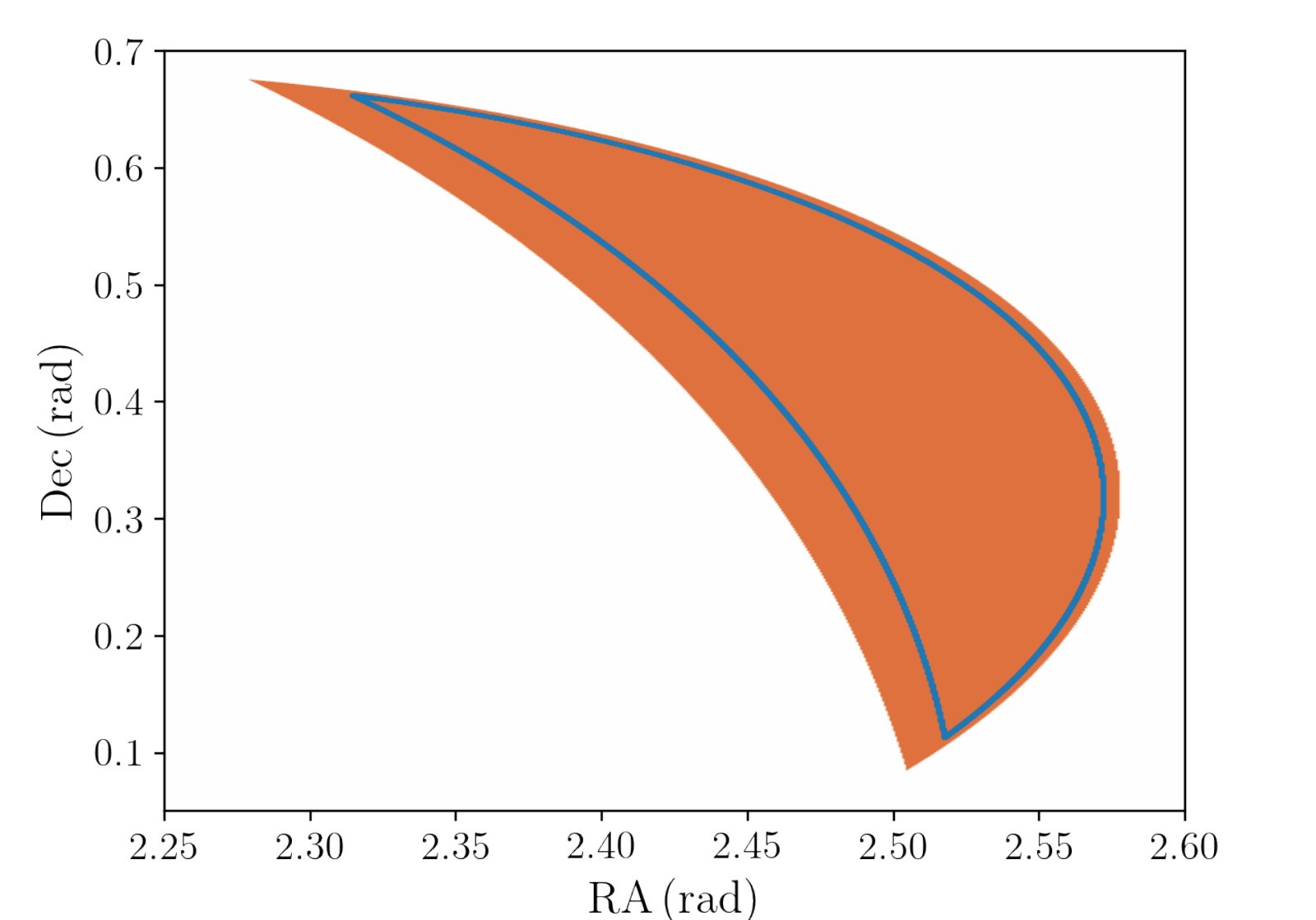}
\caption{Top: viewing angle $\delta$ as a function of the emergence angle $\theta_e$ for two different altitudes. Bottom: observable portions of the sky for one satellite position and $\alpha_{\rm off}=0\degree$, for two different emergence angles $\theta_e=14.4\degree$ (left) and $\theta_e=19.6\degree$ (right). The observable portion of the sky is the filled red area for $h=525\,{\rm km}$ and is delimited by the blue contour for $h=1000\,{\rm km}$.}\label{Fig:delta_h}
\end{figure}

The influence of the altitude is assessed by calculating the solid angle accessible for observation for one given satellite position. We consider two different altitudes $h=525\,{\rm km}$ and $h=1000\,{\rm km}$ as fiducial examples. 
The altitude of $h=525\,{\rm km}$ is taken as the lowest altitude that yields a stable orbit over a multi-year mission lifetime due to relatively insignificant atmospheric drag.
As shown before, the observable portion of the sky depends on the FOV of the detector and on the maximum emergence angle. We draw comparisons at a fixed emergence angle as this quantity is directly related to the path of the neutrino and the tau lepton through the Earth. We note that the viewing angle $\delta$ depends on the emergence angle $\theta_e$ and the altitude $h$, which is illustrated in Fig.~\ref{Fig:delta_h}. 

For a fixed maximum Earth emergence angle, $\delta$ decreases with increasing altitude and as the ``lateral'' extension of the observable portion of the sky depends on $\delta$, the solid angle accessible for observation actually decreases with increasing altitude. Some numerical estimates are illustrated in Table~\ref{Tab:Nu_results_comp}, for the various emergence angles, offset angles and altitudes considered in this paper. We also compare in Fig.~\ref{Fig:delta_h} the observable fraction of the sky projected on the celestial sphere, for one satellite position, for two different altitudes $h=525\,{\rm km}$ and $h=1000\,{\rm km}$, and for two different emergence angles $\theta_e=14.4\degree$ and $\theta_e=19.6\degree$. Interestingly, lower altitudes tend to increase the observable solid angle in the sky, for a fixed maximum Earth emergence angle. As shown in \cite{Alvarez17,Reno18}, the probability of having a tau lepton escape the Earth for $E_\tau \gsim 10$ PeV becomes relatively small for $\theta_e \gsim 20^\circ$ assuming nominal values of the cosmogenic neutrino flux, such as those in \cite{Kotera10}. Note that for a fixed delta (viewing angle away from the Earth limb) the maximum Earth emergence angle increases with increasing altitude.

\subsection{Accounting for the detection focal surface of the instrument }
\begin{figure}[ht]
\centering
\includegraphics[width=0.45\textwidth]{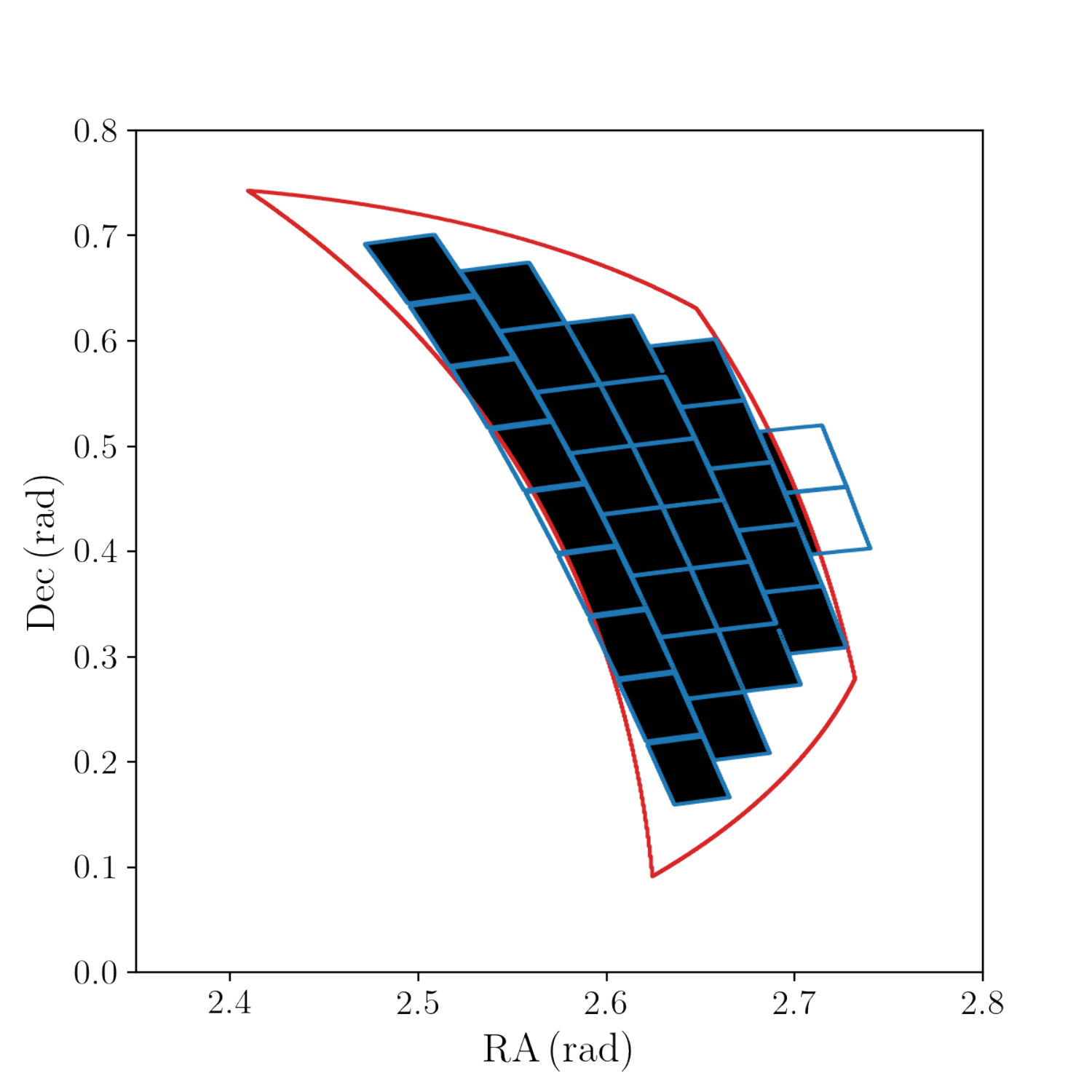} 
\includegraphics[width=0.45\textwidth]{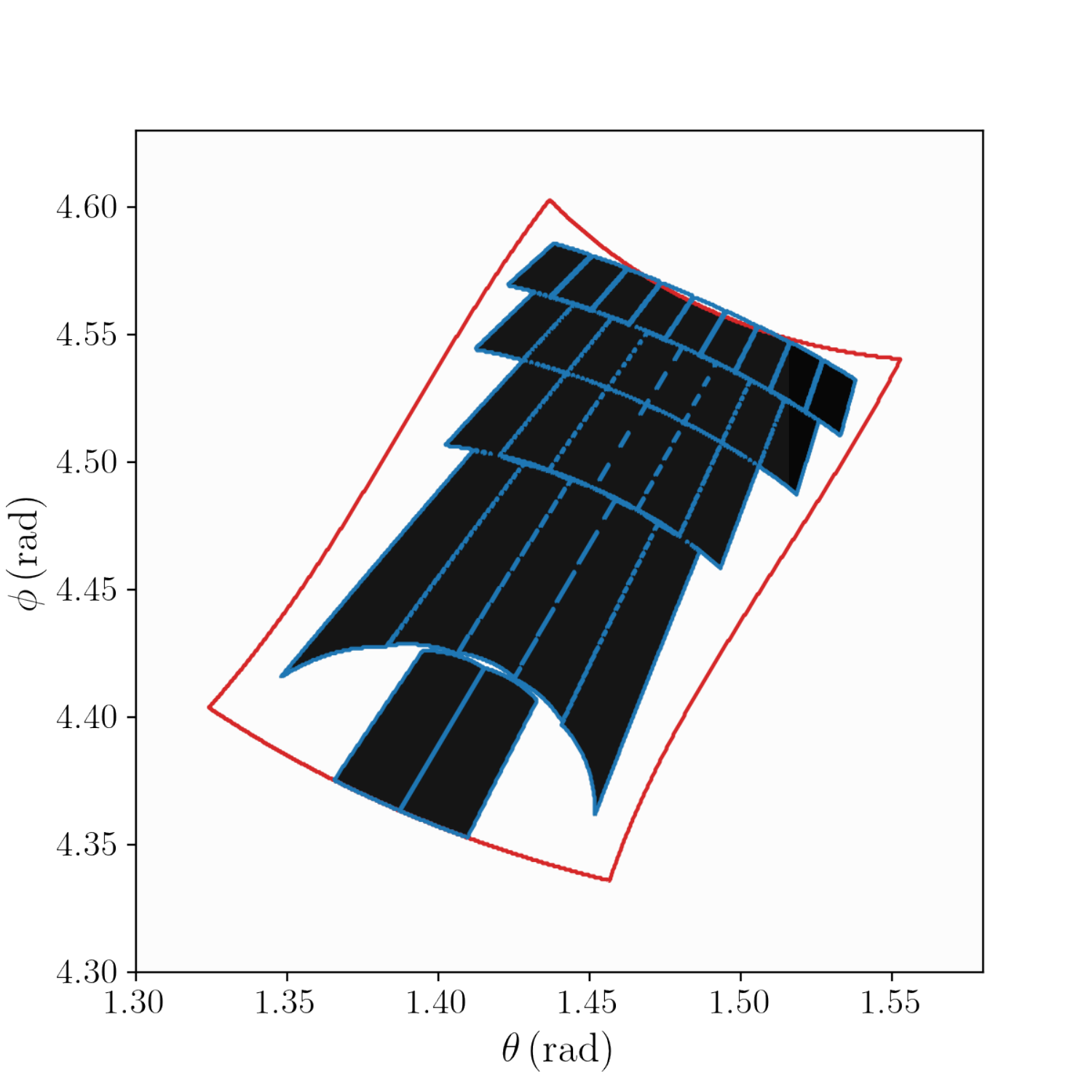} 
\caption{
Projections of the effective FOV on the celestial sphere as a function of RA and Dec (left), and on the Earth surface as a function of the spherical coordinates (right), for $h=525\,{\rm km}$, $\theta_e=19.6\degree$ and $\alpha_{\rm off}=2\degree$. We compare the geometrical FOV (red), its observable fraction (black) and the delimitation of the edges of the Cherenkov detection units (blue).}\label{Fig:focal_surf}
\end{figure}

The calculations presented above are derived from purely geometrical considerations. However, more realistic estimates should account for the dead areas on the focal surface for actual configurations of the optical Cherenkov detection units, as they usually cover only a fraction of this surface. We illustrate this for the configuration adopted within the design of the POEMMA instrument (as shown in Fig.~\ref{Fig:focal_surf_1}). In this case, only a limited part of the focal surface is dedicated to the detection of VHE neutrinos, while the remaining part is dedicated to the detection of UHECRs. The focal surface and the projection of the observable fraction of the geometrical FOV on the celestial sphere and on the Earth surface are illustrated in Fig.~\ref{Fig:focal_surf}, for $h=525\,{\rm km}$, $\theta_e=19.6\degree$ and $\alpha_{\rm off}=2\degree$.

Due to the size of the detection units, only a fraction of the geometrical FOV can lead to efficient detections. Regarding the focal surface, because of the orientation of the detector towards the limb, the geometrical FOV overlays only an outlying part of the focal surface. The small uncovered area at the edge of the focal surface is related to the offset angle of the detector. We note that in this configuration, the main part of the two upper detection units will be dedicated to background measurements two degrees above the Earth limb. This effect is illustrated in the left panel of Fig.~\ref{Fig:focal_surf}, but not on the right panel as this part of the focal surface is above the limb and thus cannot be projected on the Earth surface. On the one hand, concerning the projection on the Earth surface, we note that important distortions appear, due to projection effects of the relatively small $\theta_e$, as very limited portions of the focal surface represent a significant area on the ground. Indeed, areas are strongly stretched close to the limb. The projection on the celestial sphere, on the other hand, does not lead to significant distortions.

For $h=525\,{\rm km}$ and $\alpha_{\rm off}=2\degree$, we can compare the solid angle $\Omega$ obtained for the geometrical FOV and its observable fraction: for $\theta_e=19.6\degree$, $\Omega_{\rm obs}  = 4.9 \times 10 ^{-2}\,{\rm sr}$ ($\Omega_{\rm obs}/\Omega = 0.73$), and for $\theta_e=14.4\degree$, $\Omega_{\rm obs}  = 2.4 \times 10 ^{-2}\,{\rm sr}$ ($\Omega_{\rm obs}/\Omega = 0.71$).

\subsection{Sky exposure}

\begin{figure}[ht]
\centering
\includegraphics[width=1\textwidth]{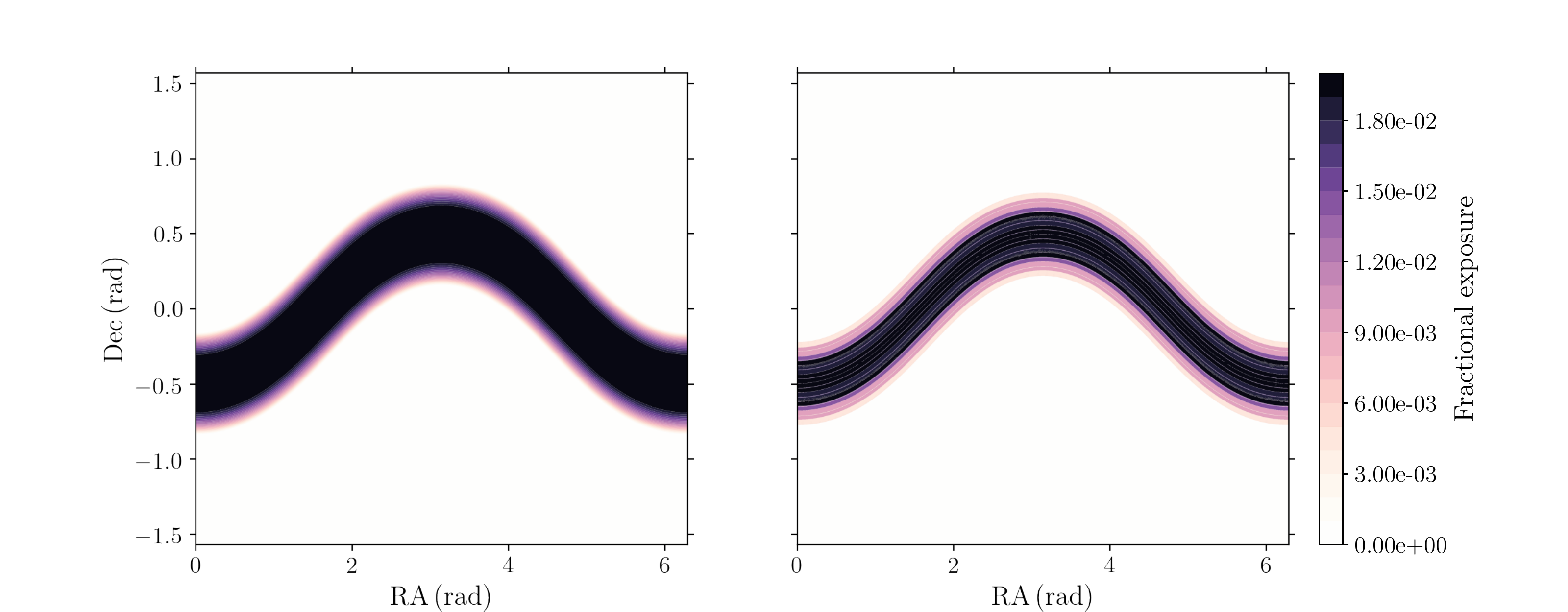}
\caption{Fractional exposure $t/P$ for $i=28.5\degree$, $\theta_e=19.6\degree$ and $h=525\,{\rm km}$ as a function of right ascension (RA) and declination (Dec), for a detector axis in the orbital plane. Comparison between the geometrical estimates when accounting or not for the effect of the detection units for the POEMMA configuration (respectively {right} and {left} panels).}\label{Fig:Nu_orb}
\end{figure}

The sky coverage for one orbit is illustrated in Fig.~\ref{Fig:Nu_orb}. We compare the two geometrical estimates accounting or not for the configuration of the detection units on the focal surface specific to POEMMA. As outlined earlier, the significant change of solid angle coverage alone can lead to significant differences in exposure calculations. The fractional exposure is the total time during which the instrument can detect neutrinos coming from each bin of size ${\rm d}\cos\Theta \,{\rm d}\Phi$, divided by the orbital period. As expected, the lateral extension of the effective FOV appears to be critical in the calculation and a fractional exposure is a bit reduced by accounting for the effect of the finite size of the active area of the Cherenkov detection units in the focal plane.
The total exposure during one orbit can be calculated by integrating $t(\Theta,\Phi)$ over the celestial sphere
\begin{equation}
t_{\rm orb} = \frac{1}{4\pi} \int \int  {\rm d}\cos\Theta \,{\rm d}\Phi\, t(\Theta,\Phi) \,.
\end{equation}
As an example, for $h=525\,{\rm km}$, $\theta_e=19.6\degree$ and $\alpha_{\rm off} = 2\degree$ we obtain the following exposure times for one orbit: $t_{\rm orb} \simeq 31\,{\rm s}$ and $t_{\rm orb} \simeq 23\,{\rm s}$ for the geometrical estimate and the realistic estimate including the effect of the detection units. We recall that for $h=525\,{\rm km}$, the period is $P \simeq 5.7 \times 10^3 \,{\rm s}$.

In our calculation we have assumed a fixed orientation of the detector, in the orbital plane and in a direction opposite to the direction of motion. This results in a limited sky coverage as a function of the declination. Indeed, during the observation time allowed by the mission, the precession around the north pole induces a shift of the sky coverage along the right ascension axis, but the declination range remains unchanged. In the next section, we show that the homogeneity of the sky coverage can be significantly increased by rotating the viewing direction out of the orbital plane.

\section{Strategy to achieve full sky coverage}\label{Sec:full_sky}

\begin{figure}[ht]
\centering
\includegraphics[width=0.49\textwidth]{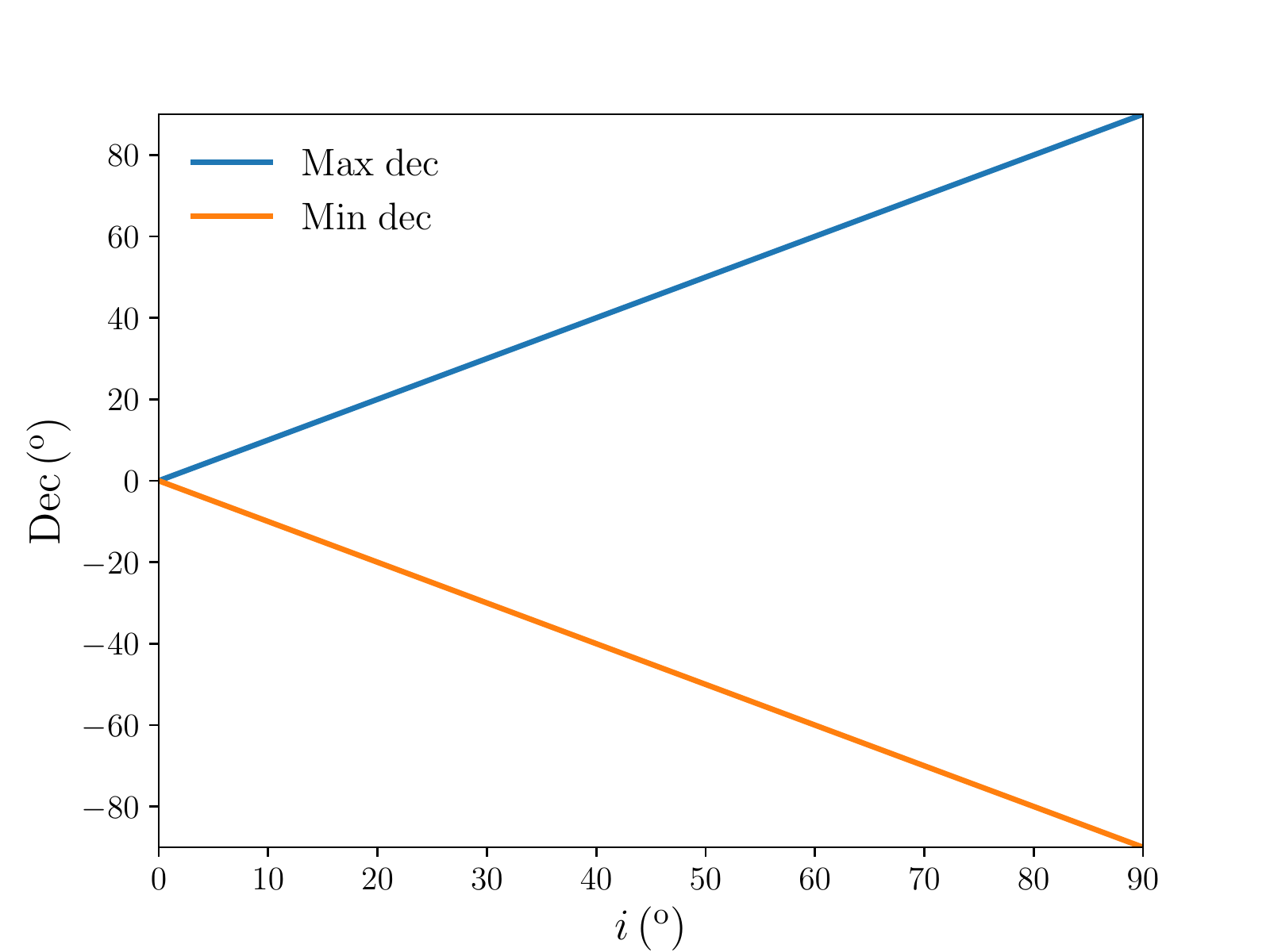}
\includegraphics[width=0.49\textwidth]{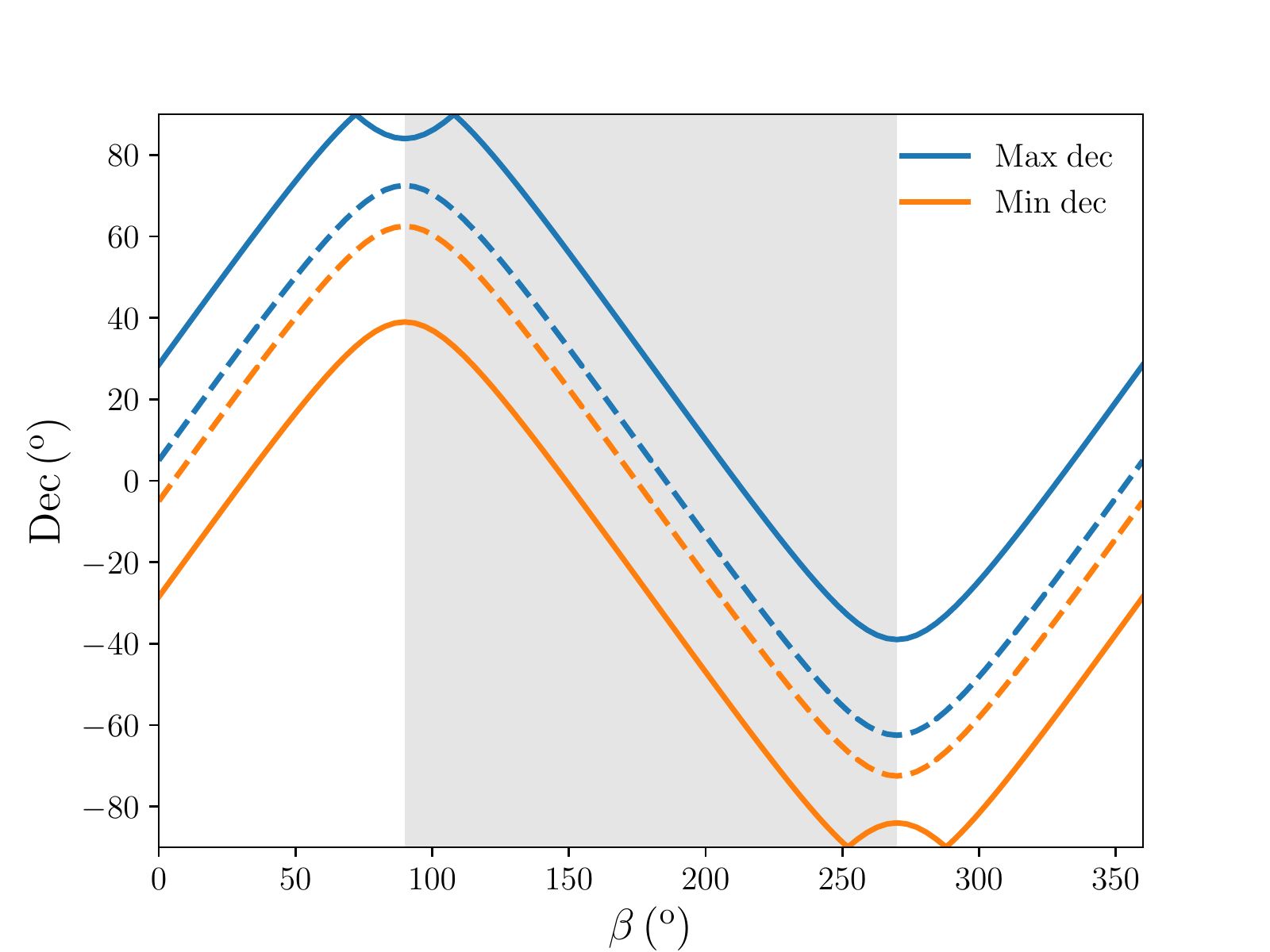}
\caption{The maximum and minimum values of declination (Dec) which can be observed (respectively blue and orange lines), as a function of the orbit inclination $i$ (left) for a detector axis in the orbital plane, and as a function of the detector inclination $\beta$ (right), for an altitude $h=525\,{\rm km}$, and for $i=28.5\degree$ (solid) and $i=5\degree$ (dashed), with the shaded area corresponding to detector inclinations that are disfavored due to being in the ram direction (see text).}\label{Fig:Dec_ibeta}
\end{figure}

For a fixed orientation of the detector in the orbital frame, only an incomplete portion of the sky is available for observation. The minimum and maximum values of declination accessible to observation do not depend on the precession angle of the orbit $j$, by spherical symmetry -- as the orbit precesses around the north pole -- but they depend on the orbit inclination $i$, as illustrated in Fig.~\ref{Fig:Dec_ibeta}.

\begin{figure}[ht]
\centering
\includegraphics[width=0.49\textwidth]{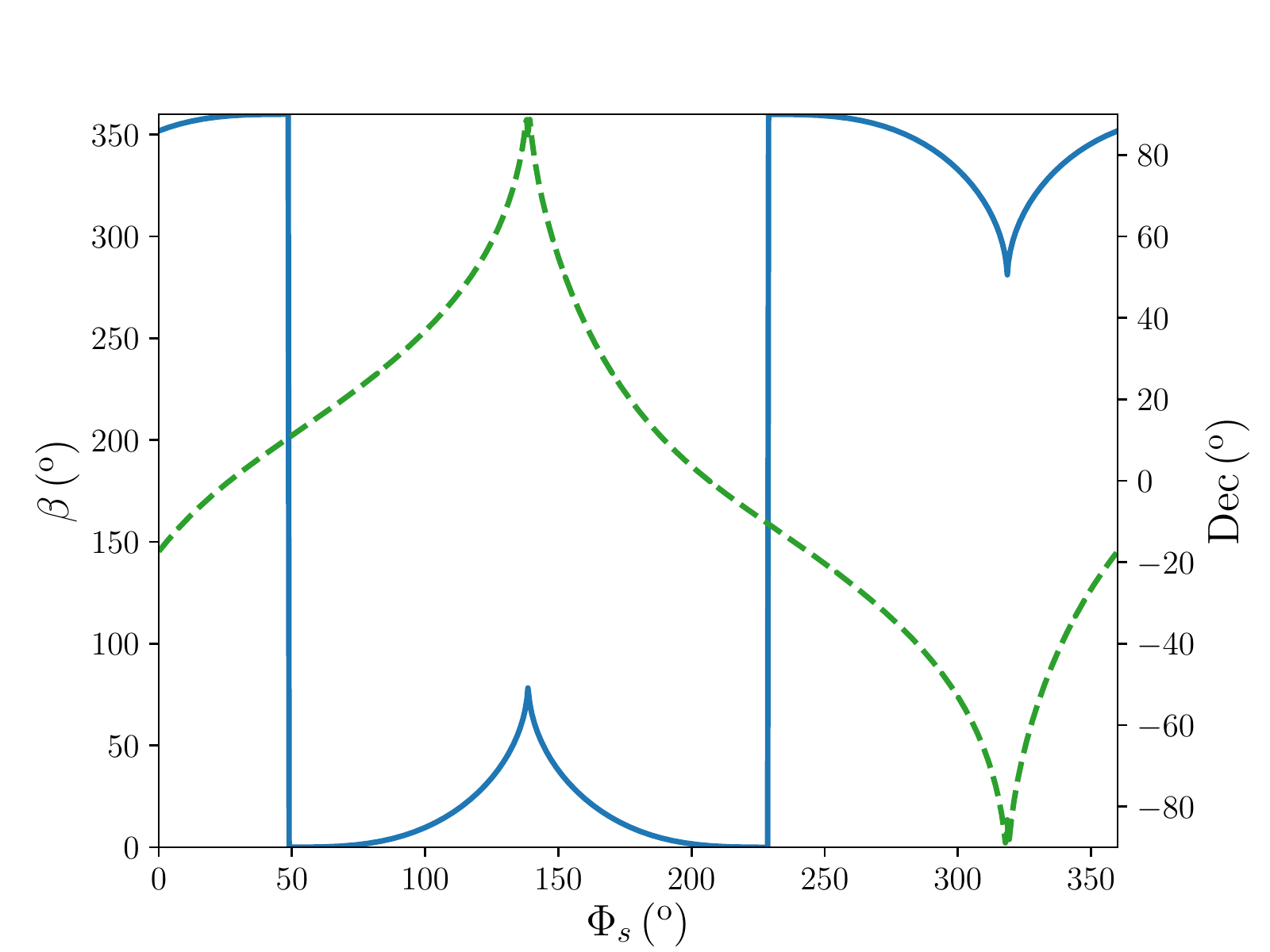}
\caption{Example of detector axis rotation during one orbit, characterized by the angle $\beta$ as a function of $\Phi_s$ (blue, solid) and the corresponding observable values of declination (Dec) as a function of $\Phi_s$ (green, dashed).}\label{Fig:Detector_pointing}
\end{figure}

The different detector orientations are characterized by the angle $\beta$:
\begin{equation}
\vec{n}_d = -[\cos(\alpha-\alpha_c+\alpha_{\rm off}) \vec{u}_{\rm sat} + \sin(\alpha-\alpha_c+\alpha_{\rm off})\cos(\beta) \vec{v}_{\rm sat} + \sin(\alpha-\alpha_c+\alpha_{\rm off})\sin(\beta)\vec{n}_{\rm orb}] \, .
\end{equation}
If $\beta=0$, the detector axis is in the orbital plane, in an opposite direction to the direction of motion. Angles around $\beta=\pi$ are not suitable if degradation of the optics from atomic oxygen is an issue. We calculate the minimum and maximum values of declination that can be observed during one orbit. To simplify the calculations, we consider that the detector is only pointing towards the limb. Moreover, we consider an orbit with $\vec{n}_{\rm orb} = \sin i \cos j \,\vec{I} + \sin i \sin j \,\vec{J} + \cos i \,\vec{K} $ such as $j=0$. We see in Fig.~\ref{Fig:Dec_ibeta} that the detector inclination, as well as the orbit inclination, have a strong influence on the portion of the sky available for observation. For the study presented here, we make the conservative assumption that the detector axis should not be oriented towards the ram direction. For $i=5\degree$, the minimum and maximum declination accessible remain respectively above $-\pi/2$ and below $\pi/2$. For $i=28.5\degree$, the values of $-\pi/2$ and $\pi/2$ can be reached for acceptable values of $\beta$ -- following our notations, the values between $\pi/2$ and $3\pi/2$ are excluded. We could therefore consider a detector axis oscillation during one orbit, characterized by a variation of the angle $\beta$ as a function of the right ascension of the satellite $\Phi_s$, allowing to cover the full declination range, and avoiding values around $\beta=\pi$. This strategy is illustrated in Fig.~\ref{Fig:Detector_pointing} for one orbit. One can also choose to rotate the detector axis over longer time scales, for instance at the end of a predefined set of orbits.

\begin{figure}[ht]
\centering
\includegraphics[width=0.49\textwidth]{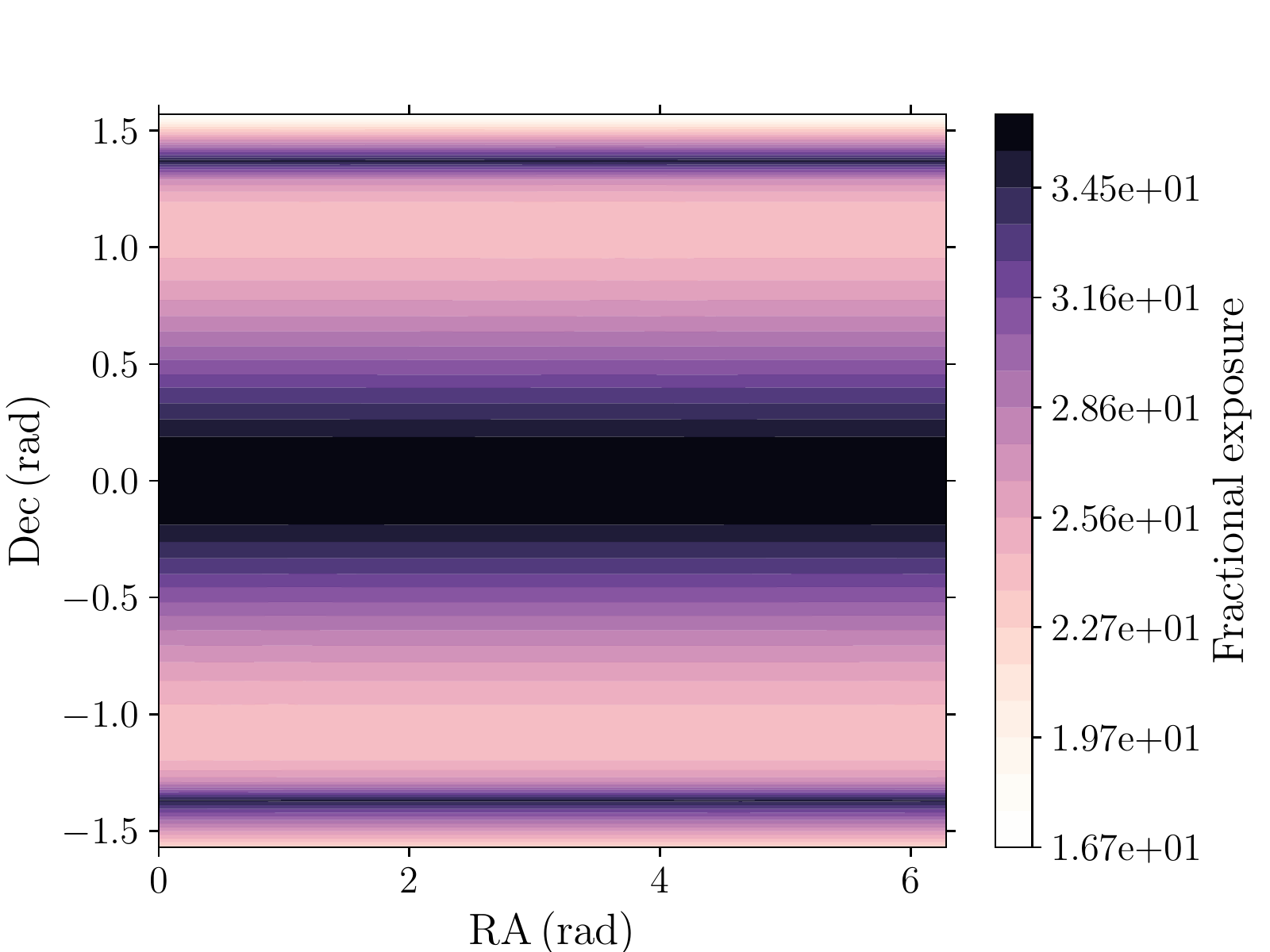}
\includegraphics[width=0.49\textwidth]{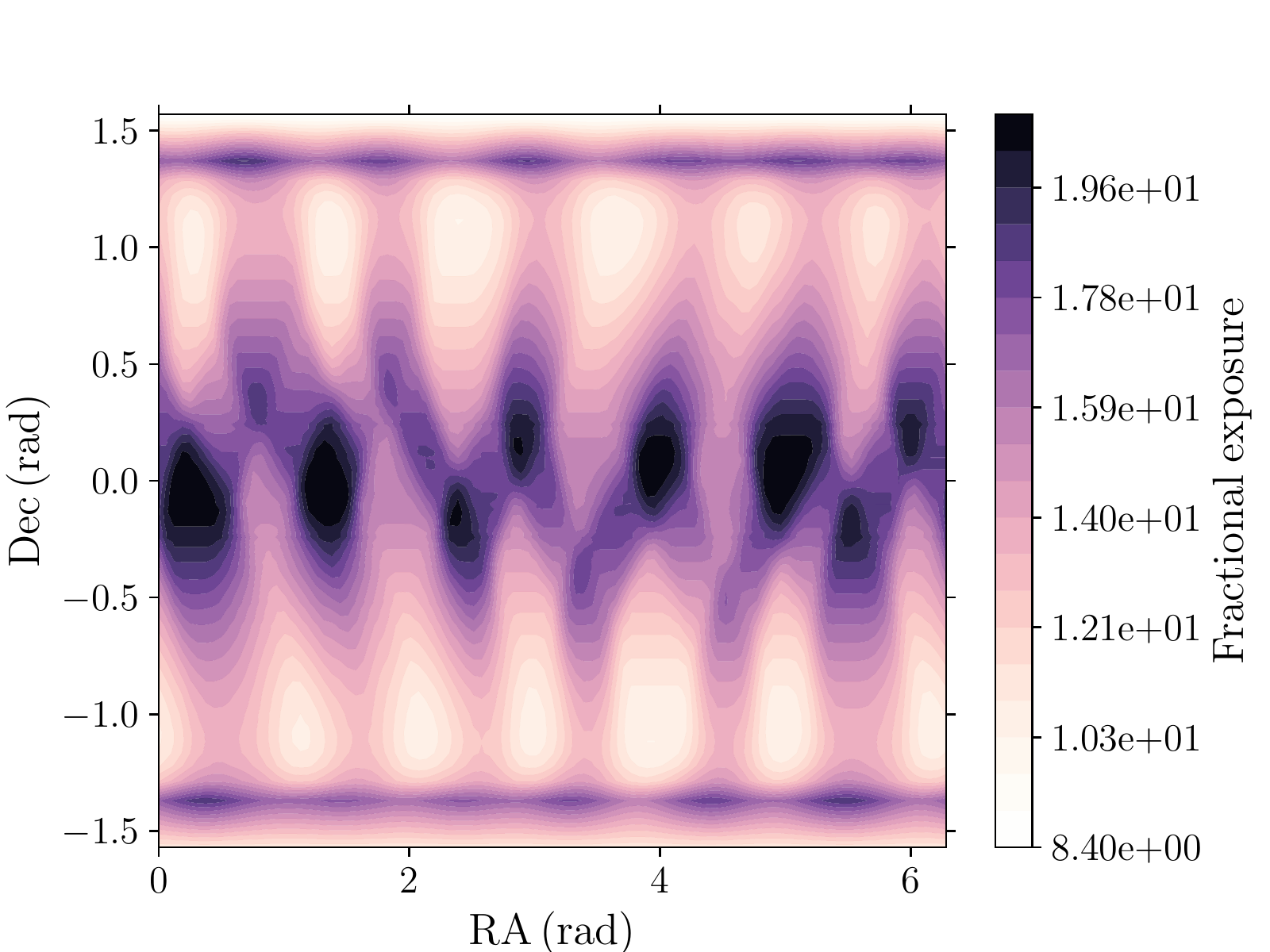}
\includegraphics[width=0.49\textwidth]{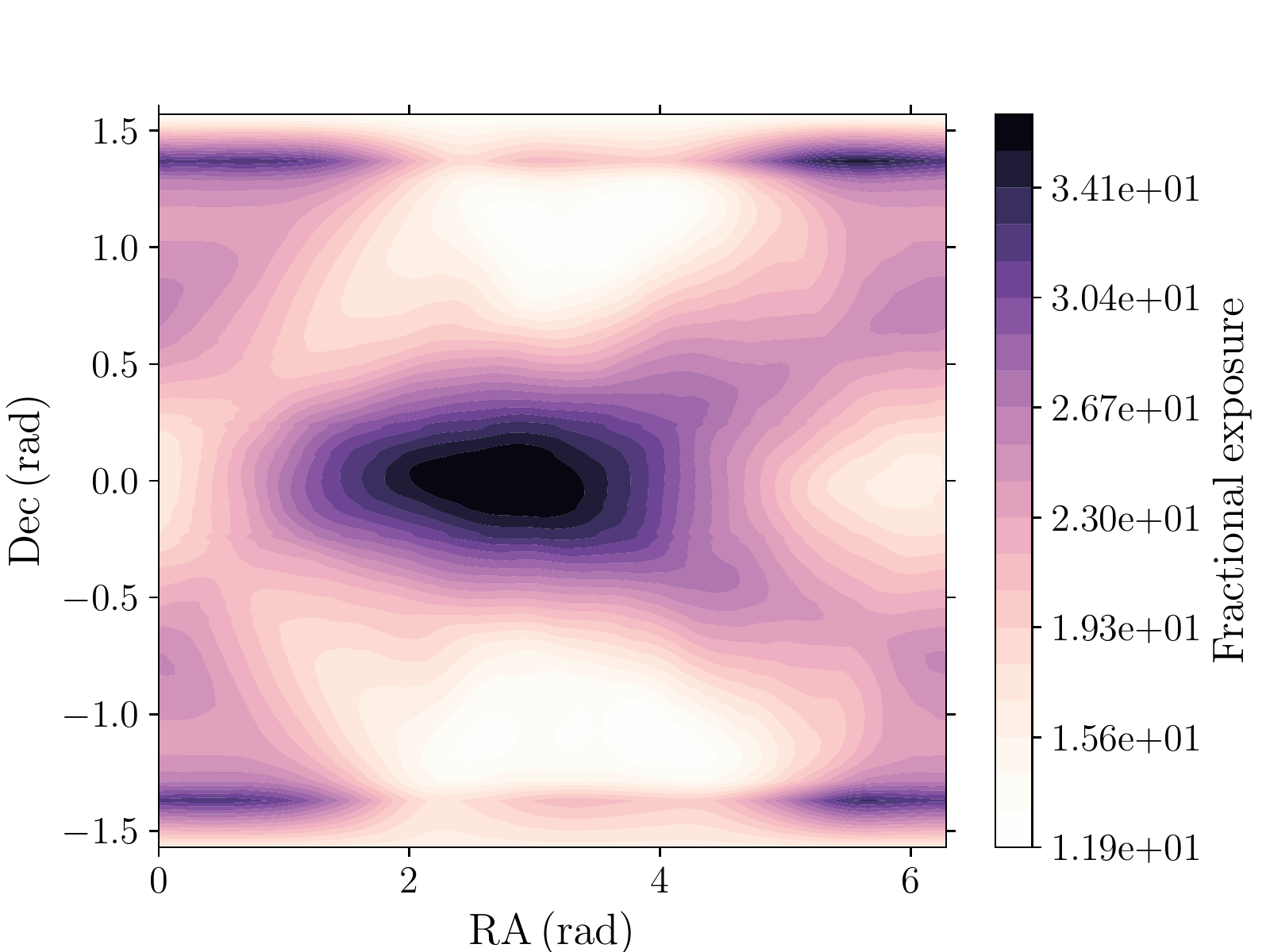}
\includegraphics[width=0.49\textwidth]{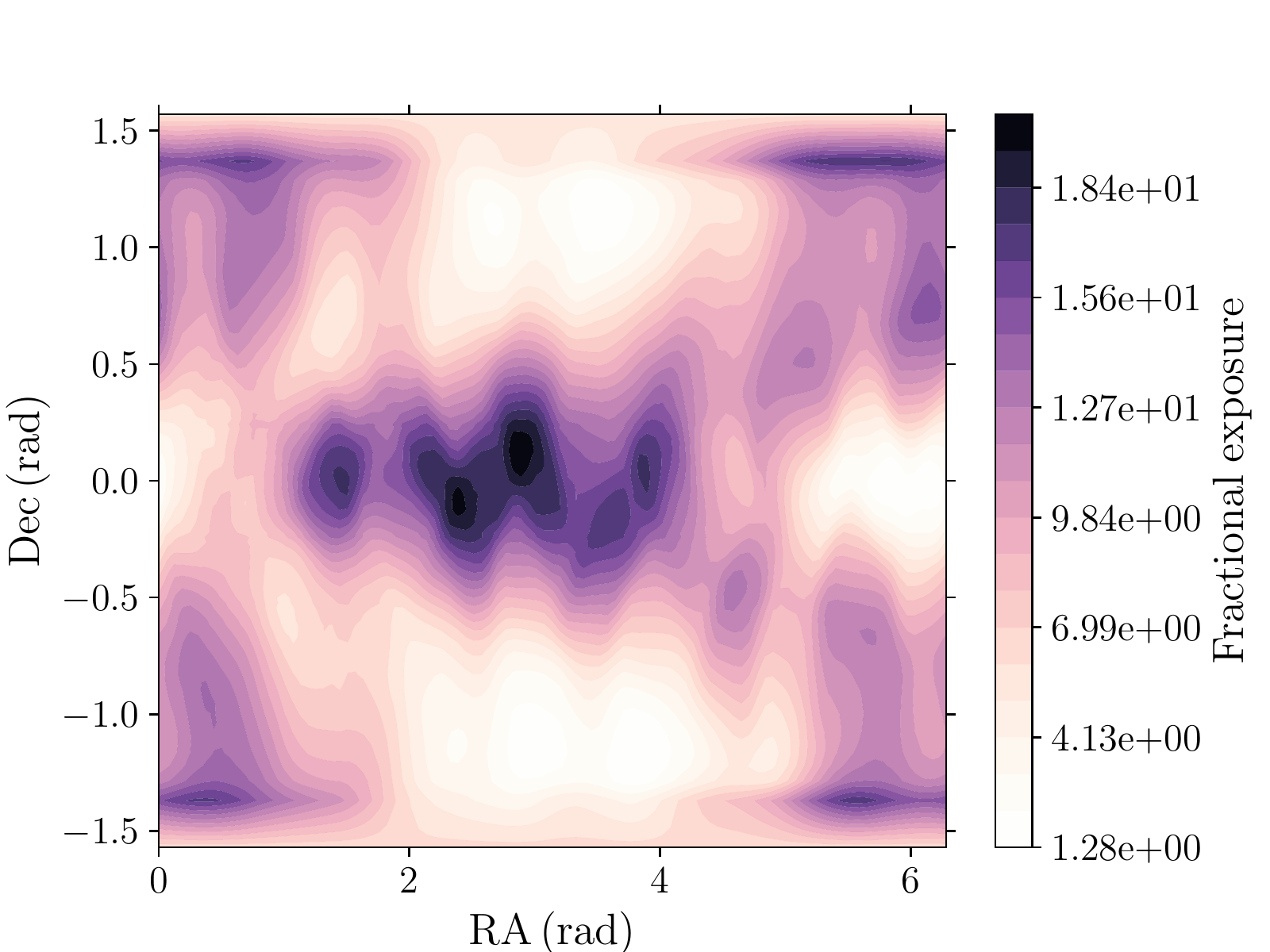}
\caption{Fractional exposure $t/P$ for $i=28.5\degree$, $h=525\,{\rm km}$ and $\theta_e=19.6\degree$ as a function of right ascension (RA) and declination (Dec), for $380$ days of observation ($7\times 2\pi / \omega_p$), with an optimized rotation of the detector axis during one orbit, for the detector layout of POEMMA. We compare the exposures obtained without the impact of the Sun and the Moon (upper left), with the impact of the Sun (upper right), with the impact of the Moon (lower left) and with the impact of the Sun and the Moon (lower right). }\label{Fig:Sky_cov_tot}
\end{figure}

Using such a strategy, with an oscillation of the detector axis during one orbit, we can calculate the sky coverage for different observation times, as illustrated in Fig.~\ref{Fig:Sky_cov_tot} for $380$ days of observation, which corresponds to seven precession periods of the satellite orbit $T_{\rm obs} = 7\times 2\pi/\omega_p$. In this calculation we consider the geometrical detector layout of POEMMA, but we do not account for the detection units as shown in Fig.~\ref{Fig:Nu_orb}. We compare the maximum sky exposure, which does not account for the impact of the Sun and the Moon in the calculation, with the exposures obtained when accounting for the impact of the Sun and the Moon. No observation can be performed if the satellite is illuminated by the Sun and if the illumination of the Earth by the Moon is too high. The Moon illumination is given by $p_{\rm moon}/100 = (1+\cos \phi)/2$, where $\phi$ is the phase angle of the Moon. We consider a maximum illumination of the Moon of $50\%$ in our calculations. We find that the Sun and the Moon have a strong impact on the sky exposure, especially on its RA dependence. The link between the precession periods of the satellite, the Sun and the Moon leads to the emergence of `hot' and `cold spots' in the sky exposure map. The exposure becomes more uniform for longer observation times. We note that our strategy allows us to obtain a uniform exposure only if the Sun and the Moon are not taken into account. In order to obtain a more uniform sky exposure, the rotation of the detector axis should be optimized by taking into account the orbits of the Sun and the Moon.

\section{Target of opportunity follow-up}\label{Sec:Nu_ToO}

\begin{figure}[ht]
\centering
\includegraphics[width=0.295\textwidth]{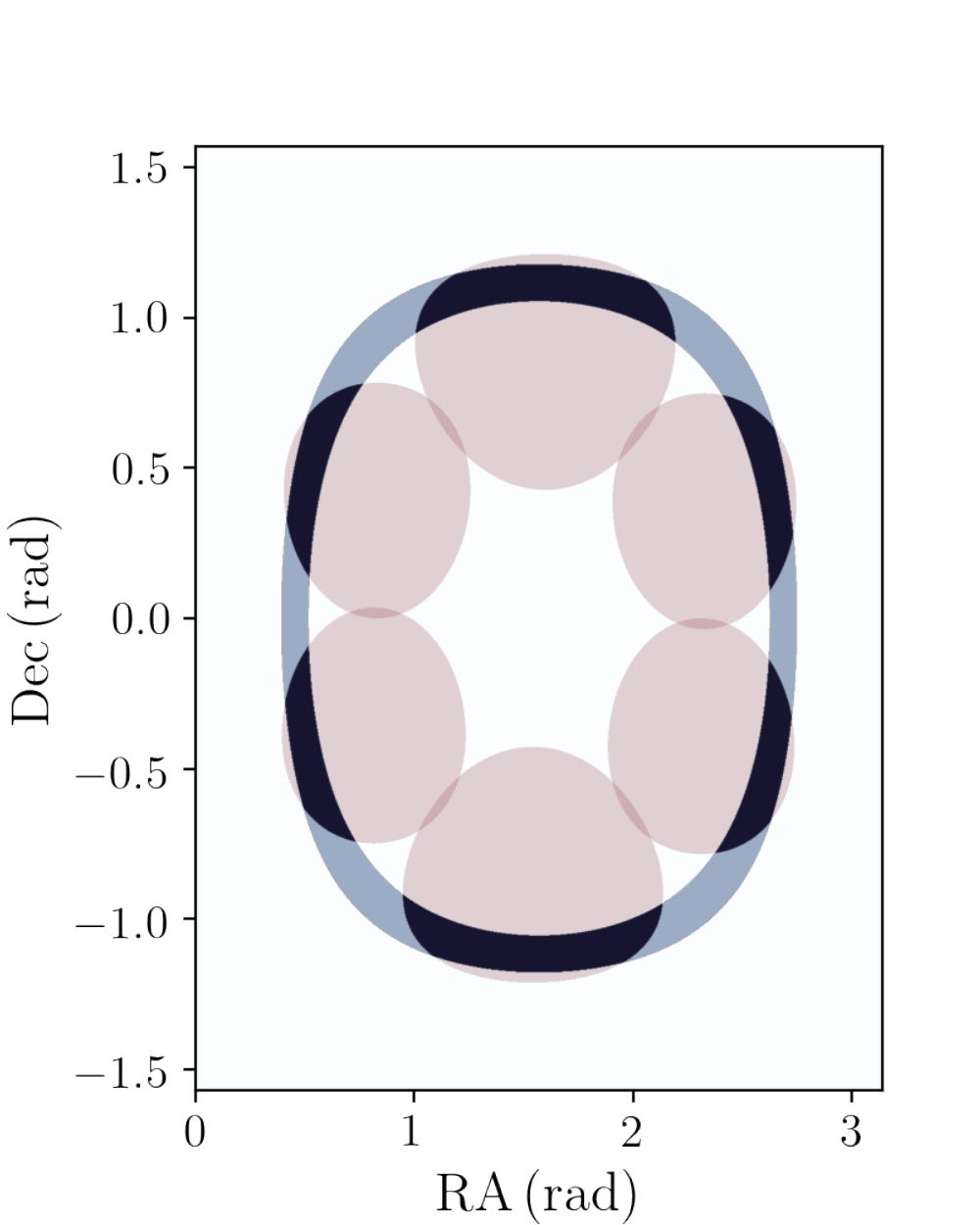}
\includegraphics[width=0.51\textwidth]{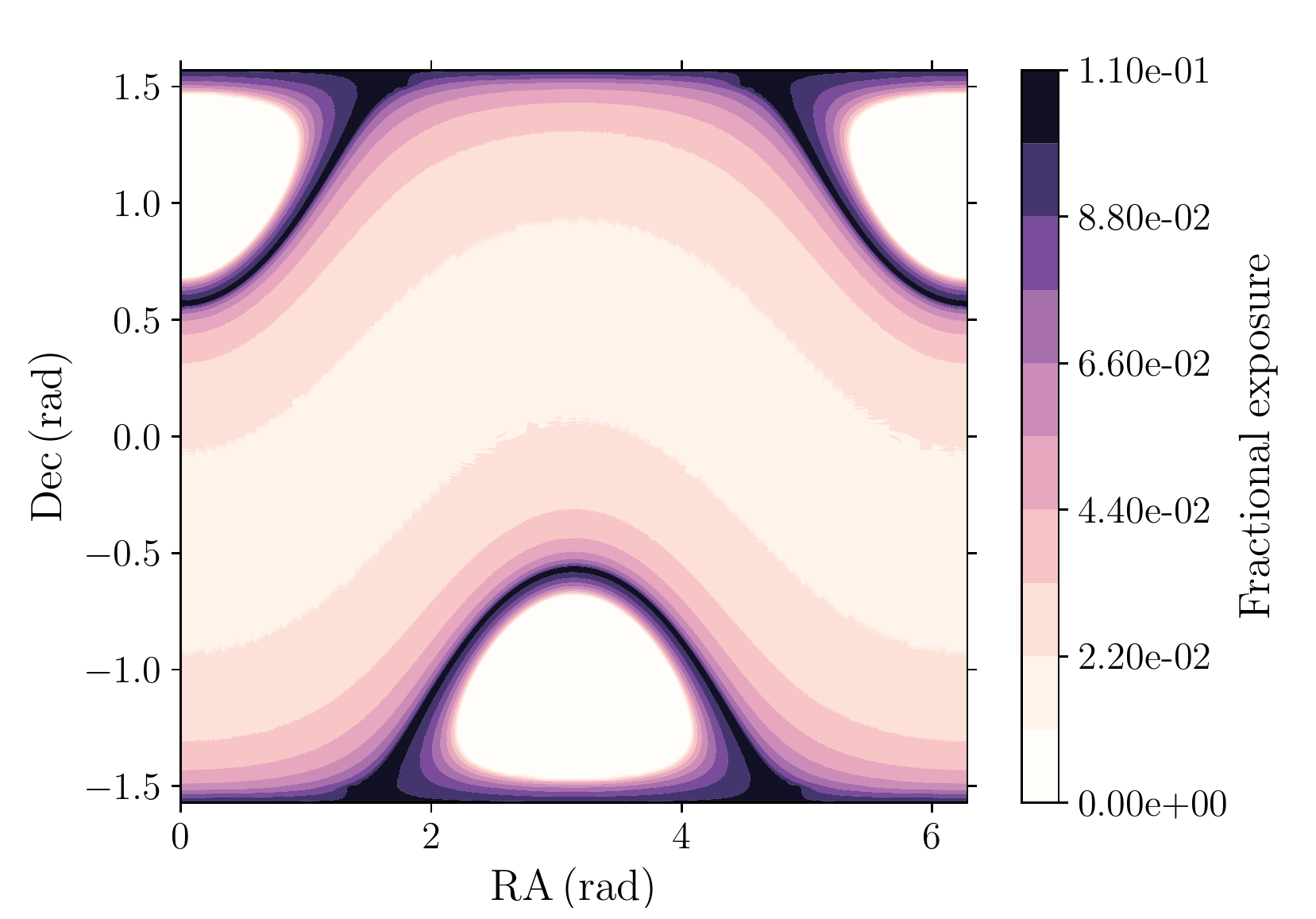}
\caption{Left: observable portion of the sky using six non-overlapping detector orientations, as a function of right ascension and declination, for $i=28.5\degree$, $h=525\,{\rm km}$, $\theta_e=19.6\degree$ and $\alpha_{\rm off} = 2\degree$. The blue band is related to the condition on the emergence angle, the red region to the condition on the FOV and the black regions to their intersection. Right: maximum fractional exposure for every direction of the sky, as a function of right ascension and declination.}\label{Fig:ToO_sky}
\end{figure}

In the case of an external alert of an interesting transient event, for instance following a localized gravitational wave detection, the neutrino space detector could enter a specific ToO observation mode, designed to maximize its exposure to a given region in the sky.  In this section we do not focus on a specific instrument, as we only use the condition of the emergence angle (see equation~\ref{Eq:cond_emangle}) in order to evaluate the observable portion of the sky during one orbit. Indeed, in the case of the POEMMA instrument, the rotation of the detector axis changes the orientation of the FOV, and the observable portion of the sky corresponds to the blue band, see Fig.~\ref{Fig:ToO_sky}, which is related to the condition on the emergence angle. As the detector axis should not be oriented towards the direction of motion, only half of this blue region is accessible to observation. By adding the exposures related to the only accessible region, we simply calculate the maximum fractional exposure that we can obtain for every direction in the sky, which is illustrated in Fig.~\ref{Fig:ToO_sky}.

We note that even if the entire declination range was accessible during one orbit (see section~\ref{Sec:full_sky}), it would not be accessible at each satellite position: in Fig.~\ref{Fig:ToO_sky} (left) we see that the condition on the emergence angle (blue band) gives minimum and maximum values of the declination accessible for observation, with a declination range of $2\alpha<\pi$ which depends on the satellite altitude. Therefore, during one orbit, some regions of the sky cannot be observed: two blind spots appear at high declinations. For $i=28.5\degree$, $h=525\,{\rm km}$, $\theta_e=19.6\degree$ and $\alpha_{\rm off}=2\degree$, about $P_{\rm obs}=91\%$ of the sky is available for observation during one orbit, given
\begin{equation}
P_{\rm obs} = \frac{1}{4\pi} \iint\displaylimits_{t(\Theta,\Phi)>0} {\rm d}\cos \Theta {\rm d}\Phi \, , 
\end{equation}
where $t(\Theta,\Phi)$ is the exposure time for any direction of the sky. As the orbit precesses slowly, with a precession period of about $54$\,days for $i=28.5\degree$ and $h=525\,{\rm km}$, some regions of the sky will not be accessible for short duration transient phenomena. 

\begin{figure}[ht]
\centering
\includegraphics[width=0.49\textwidth]{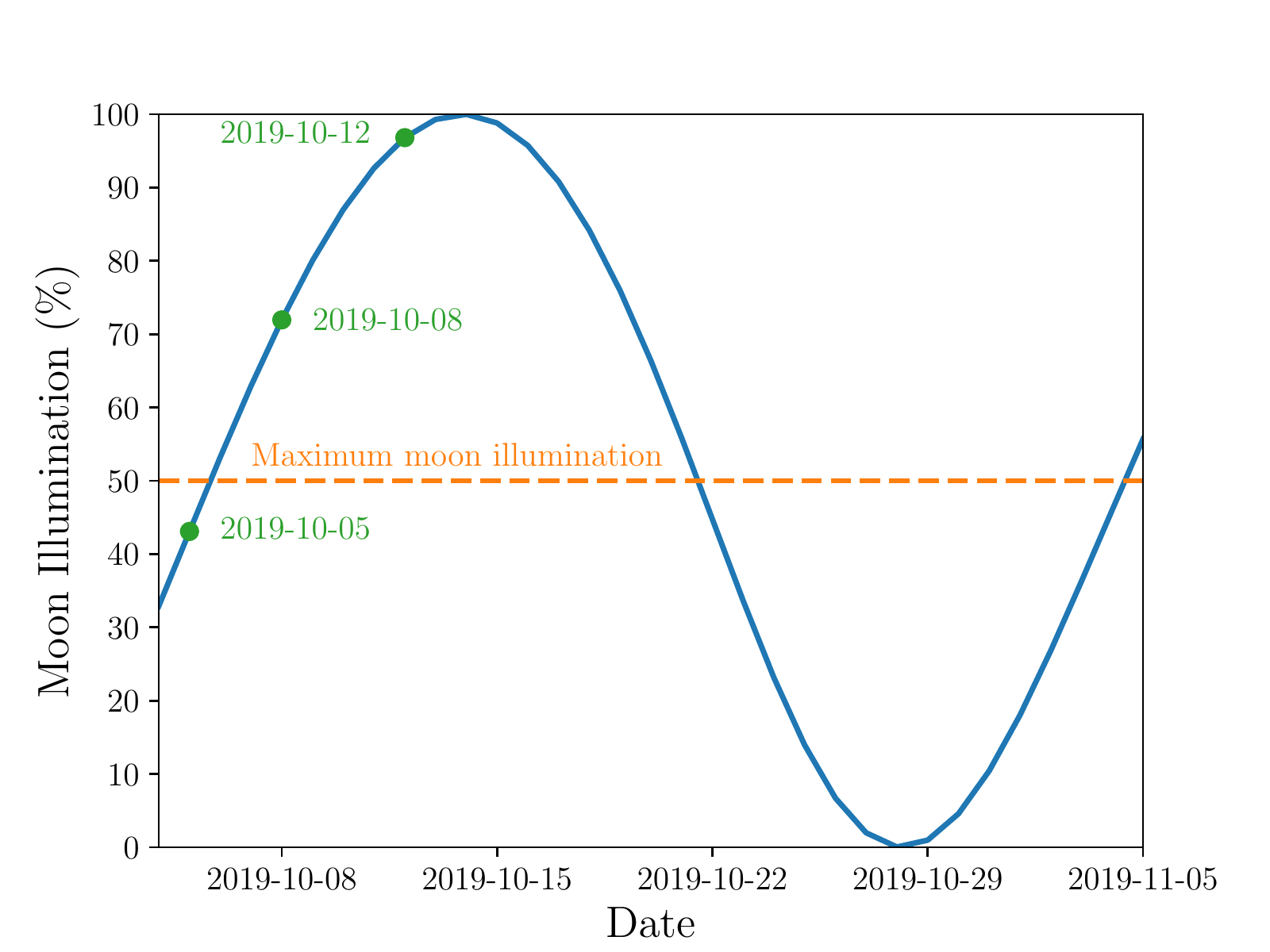}
\includegraphics[width=\textwidth]{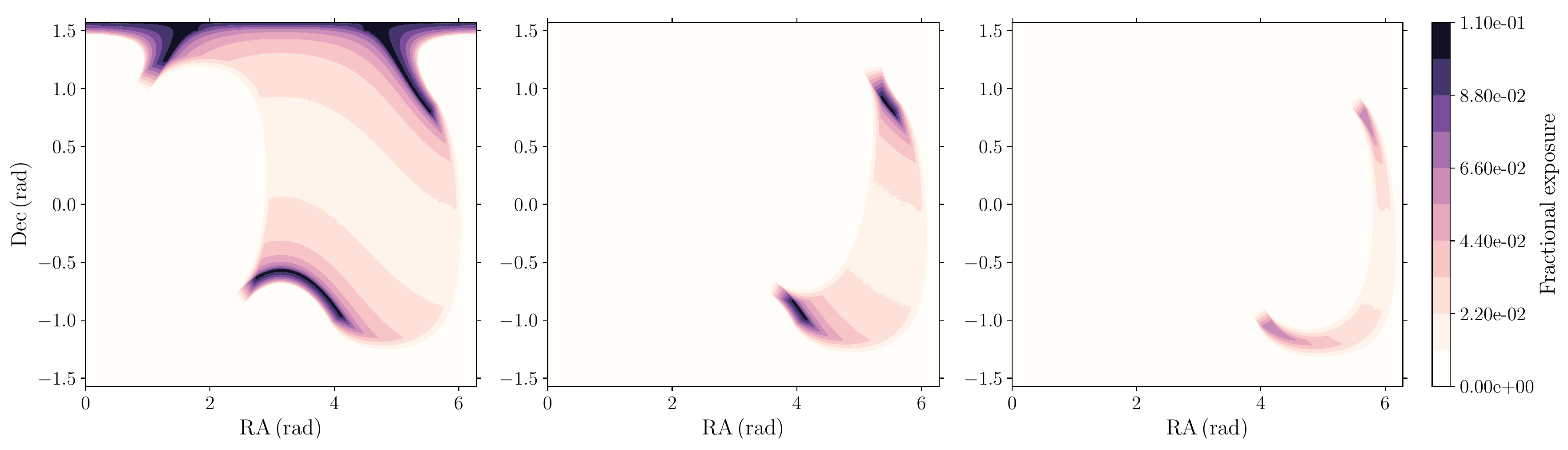}
\caption{Top: Moon illumination during one lunar cycle (blue line), fiducial maximum Moon illumination (orange dashed line) and three dates used as examples (green points). Bottom: maximum fractional exposure for every direction of the sky, as a function of right ascension and declination, for the three dates (left: 2019-10-5, middle: 2019-10-08, right: 2019-10-12), including the impact of the Sun and the Moon.}\label{Fig:ToO_sky_moon}
\end{figure}

As emphasized previously, the Sun and the Moon illuminations have a strong impact on the observations and cannot be neglected. During about half of the orbit, the satellite is illuminated by the Sun, and thus about $49\%$ of the sky is available for observation during one orbit, for the parameters given above. The regions of the sky occulted by the Sun will remain inaccessible for up to half a year. The ability to carry out timely follow-up observations of a ToO will be extremely dependent on its relative position with respect to the Sun. The Moon illumination varies throughout the lunar cycle. If the illumination is too high, observations can only be performed when the Sun and the Moon are hidden by the Earth. The observable portion of the sky during one orbit is illustrated in Fig.~\ref{Fig:ToO_sky_moon} for several dates, and thus various Moon illuminations. For illustration purposes, we consider that observations can be performed for a Moon illumination below $50\%$. We see that the observable portion of the sky can be strongly reduced during the periods of highest illumination.

We should add that our calculation does not take into account considerations related to the cloud coverage, which could hinder further the ability of an optical instrument, such as POEMMA, to search for VHE tau neutrinos from a flaring source for example, unless one considers that useful Cherenkov signal is produced in cloud-free regions or above the cloud height.

\section{Conclusion}

In this paper we show that space-based experiments that use the optical Cherenkov signal from tau-induced EASs from VHE Earth-skimming tau neutrinos can achieve full sky coverage assuming at least a year observation time,  using POEMMA as an example. These results highlight one benefit of space-based observations versus ground-based observations at a fixed geographic location. Our result is based on geometrical calculations of the FOV that also details the dependence of the sky exposure and sky coverage for various configurations, including different assumptions of the satellite altitude, maximum Earth emergence angle of the tau lepton, offset angle when viewing the Earth limb, orientation of the director axis relative to the orbit trajectory, and constraints imposed by the finite active area of photo-detectors in the focal plane of the Cherenkov telescope.  Calculations of the Earth emerging tau lepton spectrum induced from tau neutrino interactions in the Earth \cite{Alvarez17, Reno18} show that for a given tau energy threshold, the tau flux becomes significantly reduced at a specific Earth emergence angle, thus effectively setting a maximum useful Earth emergence angle. Once this is fixed, our calculations show, perhaps counter-intuitively, that the sky exposure decreases for an increasing satellite altitude.  This is due to the satellite viewing angle away from the horizon decreasing as the altitude increases when the maximum Earth emergence angle is fixed.

Without accounting for the impact of the Sun, for a maximum emergence angle $\theta_e=19.6\degree$ and a detector offset angle $\alpha_{\rm off}=2\degree$ above the Earth limb, about $91\%$ of the sky is available for observation during one orbit for a satellite altitude of $h=525\,{\rm km}$ against $87\%$ for $h=1000\,{\rm km}$. With the impact of the Sun, we obtain $49\%$ for $h=525\,{\rm km}$ and $45\%$ for $h=1000\,{\rm km}$ for one orbit. A lower altitude increased therefore the sky coverage and could also be advantageous for photon collection. As expected, the maximum Earth emergence angle strongly affects the exposure, and as this maximum angle depends on the incoming neutrino spectrum, one should observe a distribution in energy as a function of the emergence angle that in turn influences the sensitivity of the instrument.

The rotation of the detector axis during the observation time is critical in order to cover the entire declination range and thus to achieve full sky coverage. It should be noted that the POEMMA satellites can slew 90 degrees within several minutes. By rotating the detector axis during one orbit or over longer time scales, one can scan the entire declination range for $h=525\,{\rm km}$, but the highest declinations cannot be reached for $h=1000\,{\rm km}$. Without accounting for the impact of the Sun and the Moon and with an appropriate detector orientation change strategy, about one precession period is therefore needed to achieve full sky coverage. However, if we account for the presence of the Sun, some regions of the sky are simply not available for observations for up to six months. Follow-up Target-of-Opportunity observations of a transient source are therefore strongly constrained by its relative position with respect to the Sun. This is a limitation of using the optical Cherenkov from tau-induced EAS. It should be noted that measuring the radio emission from upward-moving EAS would not have the solar or lunar constraints that are required for optical Cherenkov EAS measurements, although radio detection from space presents its own set of unique challenges.

\appendix

\section{Acknowledgement}

We thank the POEMMA collaboration for several discussions related to this work. We thank the support of the FACCTS program and the NASA Awards 80NSSC18K0246, NNX17AJ82G, and NNX13AH54G at the University of Chicago, and the NASA Awards 80NSSC18K0477 and NNX13AH55G at the Colorado School of Mines and under proposal 16-APROBES16-0023 at NASA/GSFC. Claire Gu\'epin is supported by a fellowship from the CFM Foundation for Research and by the Labex ILP (reference ANR-10-LABX-63, ANR-11-IDEX-0004-02).

\bibliography{Paper_SkyCov}

\end{document}